\documentstyle[psfig,12pt,citesort,fullpage]{article}
\def\myendproof{{\ \vbox{\hrule\hbox{%
   \vrule height1.3ex\hskip0.8ex\vrule}\hrule }}\par}
\newtheorem{theorem}{Theorem}[section]
\newtheorem{lemma}[theorem]{Lemma}

\newenvironment{proof}{{\it Proof. }}{\myendproof}
\newcommand{\code}[1]{{\rm code}[#1]}
\newcommand{\select}[1]{{\rm select}({#1})}
\newcommand{\rank}[1]{{\rm rank}({#1})}
\newcommand{\match}[1]{{\rm match}({#1})}

\newcommand{\enclose}[1]{{\rm enclose}({#1})} 

\newcommand{\enclosek}[2]{{\rm enclose}_{#1}({#2})}
\newcommand{\firstk}[2]{{\rm first}_{#1}({#2})}
\newcommand{\lastk}[2]{{\rm last}_{#1}({#2})}
\newcommand{\setof}[1]{\{{#1}\}}

\newcommand{\floor}[1]{{\lfloor{#1}\rfloor}}
\newcommand{\leftp}[1]{{\mbox{\tt (}_{#1}}}
\newcommand{\rightp}[1]{{\mbox{\tt )}_{#1}}}
\newcommand{\leftb}[1]{{\mbox{\tt [}_{#1}}}
\newcommand{\rightb}[1]{{\mbox{\tt ]}_{#1}}}
\newcommand{\aux}{\alpha}
\newcommand{\Xomit}[1]{}
\newtheorem{fact}{Fact}

\title{Compact Encodings of Planar Graphs \\ via Canonical Orderings
and Multiple Parentheses\thanks{An extended abstract appeared in {\em
Proceedings of the 25th International Colloquium on Automata,
Languages, and Programming}, pages 118--129, 1998.}}

\author{ 
Richie  Chih-Nan Chuang\thanks{Department of Computer Science and
Information  Engineering, National Chung-Cheng University, Chia-Yi 621,
Taiwan, ROC.}
\and
Ashim Garg\thanks{Department of Computer Science and Engineering,
State University of New York at Buffalo, Buffalo, NY 14260, USA.
Email: agarg@cse.buffalo.edu.}
\and
Xin He\thanks{Department of Computer Science and Engineering, State
University of New York at Buffalo, Buffalo, NY 14260, USA.  Email:
xinhe@cse.buffalo.edu.  Research supported in part by NSF Grant
CCR-9205982.}
\and
Ming-Yang  Kao\thanks{Department of Computer Science, Yale University, 
New Haven,  CT 06250, USA.  Email: kao-ming-yang@cs.yale.edu.  Research 
supported in part by NSF  Grant CCR-9531028.}
\and
Hsueh-I  Lu\thanks{Department of Computer Science and Information
Engineering,  National Chung-Cheng University, Chia-Yi  621,
Taiwan,  ROC. Email: hil@cs.ccu.edu.tw. 
Research supported in part by NSC Grant NSC-88-2213-E-194-022.}}

\newcommand{\kfrac}[0]{(5+\frac{1}{k})}

\begin{document}
\maketitle

\begin{abstract} 
Let $G$ be a plane graph of $n$ nodes, $m$ edges, $f$ faces, and no
self-loop. $G$ need not be connected or simple (i.e., free of multiple
edges).  We give three sets of coding schemes for $G$ which all take
$O(m+n)$ time for encoding and decoding. Our schemes employ new
properties of canonical orderings for planar graphs and new techniques
of processing strings of multiple types of parentheses.  

For applications that need to determine in $O(1)$ time the adjacency
of two nodes and the degree of a node, we use
$2m+{\kfrac}n+o(m+n)$ bits for any constant $k>0$
while the best previous bound by Munro and Raman is $2m+8n +
o(m+n)$. If $G$ is triconnected or triangulated, our bit count
decreases to $2m+3n + o(m+n)$ or $2m+2n + o(m+n)$, respectively.  If
$G$ is simple, our bit count is $\frac{5}{3}m+{\kfrac}n+o(n)$ for
any constant $k>0$. Thus, if a simple $G$ is also triconnected or
triangulated, then $2m+2n+o(n)$ or $2m+n+o(n)$ bits suffice,
respectively.

If only adjacency queries are supported, the bit counts for a general
$G$ and a simple $G$ become $2m+\frac{14}{3}n+o(m+n)$ and
$\frac{4}{3}m+5n+o(n)$, respectively.

If we only need to reconstruct $G$ from its code, a simple and
triconnected $G$ uses $\frac{3\log_2 3}{2}m+O(1)\approx 2.38m+O(1)$
bits while the best previous bound by He, Kao, and Lu is $2.84m$. 
% THE FOLLOWING TWO LINES IN THE ABSTRACT ARE COMMENTED OUT.
% If $G$ is simple and triangulated, we use only $\frac{4}{3}m+O(1)$ bits,
% also less than the best previous bound of $1.53m$ by Keeler and Westbrook.
\end{abstract}

%\begin{keywords}
%data compression, graph encoding, canonical ordering, planar graphs,
%triconnected graphs, triangulations
%\end{keywords}

%\begin{AMSMOS}
%05C30, 05C78, 05C85, 68R10
%\end{AMSMOS}

% 05C30 Enumeration of graphs and maps 
% 05C78 Graph labeling (graceful graphs, bandwidth, etc.) 
% 05C85 Graph algorithms
% 68R10 Graph theory

\section{Introduction}
This paper investigates the problem of encoding a given graph $G$ into a
binary string $S$ with the requirement that $S$ can be decoded to
reconstruct $G$. This problem has been extensively studied with three
objectives: (1) minimizing the length of $S$, (2) minimizing the time
needed to compute and decode $S$, and (3) supporting queries
efficiently.

As these objectives are often in conflict, a number of coding schemes
with different trade-offs have been proposed. The standard
adjacency-list encoding of a graph is widely useful but requires
$2m\lceil \log n\rceil$\footnote{All logarithms are to base 2.}  bits
where $m$ and $n$ are the numbers of edges and nodes, respectively. A
folklore scheme uses $2n$ bits to encode a rooted $n$-node tree into a
string of $n$ pairs of balanced parentheses.  Since the total number
of such trees is at least
$\frac{1}{2(n-1)}{\cdot}\frac{(2n-2)!}{(n-1)!(n-1)!}$, the minimum
number of bits needed to differentiate these trees is the log of this
quantity, which is $2n-o(n)$ by Stirling's approximation
formula. Thus, two bits per edge up to an additive $o(1)$ term is an
information-theoretic tight bound for encoding rooted trees.  The
rooted trees are the only nontrivial graph family with a known
polynomial-time coding scheme whose code length matches the
information-theoretic bound.

For certain graph families, Kannan, Naor and Rudich~\cite{KNR92} gave
schemes that encode each node with $O(\log n)$ bits and support
$O(\log n)$-time testing of adjacency between two nodes. For dense
graphs and complement graphs, Kao, Occhiogrosso, and Teng~\cite{kaot93.joa} 
devised
two compressed representations from adjacency lists to speed up basic
graph techniques. Galperin and Wigderson \cite{GW83} and Papadimitriou
and Yannakakis~\cite{PH86.encode} investigated complexity issues
arising from encoding a graph by a small circuit that computes its
adjacency matrix. For labeled planar graphs, Itai and Rodeh
\cite{IR82} gave an encoding of $\frac{3}{2} n \log n + O(n)$ bits.
For unlabeled general graphs, Naor \cite{naor90} gave an encoding of
${\frac{1}{2}}n^2-n\log{n}+ O(n)$ bits.

Let $G$ be a plane graph with $n$ nodes, $m$ edges, $f$ faces, and no
self-loop. $G$ need not be connected or simple (i.e., free of multiple
edges).  We give coding schemes for $G$ which all take $O(m+n)$ time
for encoding and decoding.  The bit counts of our schemes depend on
the level of required query support and the structure of the encoded
family of graphs. In particular, whether multiple edges (or
self-loops) are permitted plays a significant role.

For applications that require support of certain queries, Jacobson
\cite{Jacobson89} gave an $\Theta(n)$-bit encoding for a connected and
simple planar graph $G$ that supports traversal in $\Theta(\log{n})$
time per node visited. Munro and Raman~\cite{MR97} recently improved
this result and gave schemes to encode binary trees, rooted ordered
trees and planar graphs. For a general planar $G$, they used
$2m+8n+o(m+n)$ bits while supporting adjacency and degree queries in
$O(1)$ time. We reduce this bit count to
$2m+{\kfrac}n+o(m+n)$ for any constant $k>0$ with
the same query support. If $G$ is triconnected or triangulated, our
bit count further decreases to $2m+3n+o(m+n)$ or $2m+2n+o(m+n)$,
respectively.
With the same query support, we can encode a simple $G$ using only
$\frac{5}{3}m+{\kfrac}n+o(n)$ bits for any constant
$k>0$. As a corollary, if
a simple $G$ is also triconnected or triangulated, the bit count is
$2m+2n+o(n)$ or $2m+n+o(n)$, respectively.

If only $O(1)$-time adjacency queries are supported, our bit counts
for a general $G$ and a simple $G$ become $2m+\frac{14}{3}n+o(m+n)$
and $\frac{4}{3}m+5n+o(n)$, respectively. All our schemes mentioned so
far as well as that of \cite{MR97} can be modified to accommodate
self-loops with $n$ additional bits.

If we only need to reconstruct $G$ with no query support, the code
length can be substantially shortened. For this case, Tur\'{a}n
\cite{turan84} used $4m$ bits for $G$ that may have self-loops; this
bound was improved by Keeler and Westbrook \cite{KW:encodings} to
$3.58m$ bits.  They also gave coding schemes for several important
families of planar graphs. In particular, they used $1.53m$ bits for a
triangulated simple $G$, and $3m$ bits for a connected $G$ free of
self-loops and degree-one nodes. 
For a simple triangulated $G$, He, Kao, and Lu~\cite{hkaol.coding_one}
improved the count to $\frac{4}{3}m+O(1)$.  Tutte~\cite{Tutte62} gave
an information-theoretic tight bound of roughly $1.08m$ bits for a
triangulated $G$. For a simple $G$ that is free of self-loops,
triconnected and thus free of degree-one nodes, He, Kao, and
Lu~\cite{hkaol.coding_one} improved the count to at most $2.84m$.  We further
improve the bit count to at most $\frac{3\log
3}{2}m+O(1)$. Figure~\ref{fig_summary} summarizes our results and
compares them with previous ones.

\begin{figure*}[t]\label{fig_summary}
\begin{center}
\begin{tabular}{|c||c|c|c|c|c|c|}  \hline
   &\multicolumn{2}{c|}{adjacency and degree}
  &\multicolumn{2}{c|}{adjacency}
  &\multicolumn{2}{c|}{no query}\\
\cline{2-7}
 &\cite{MR97}&ours&old&ours&\cite{KW:encodings,hkaol.coding_one}&ours\\
\hline
\hline
  \begin{tabular}{c}
  self-loops
  \end{tabular}
  &&&&&$3.58m$&\\
\hline
  general &
  $2m+8n$&$2m+{\kfrac}n$& &$2m+\frac{14}{3}n$& & \\
\hline
  \begin{tabular}{c}
  simple 
  \end{tabular}
  &&$\frac{5}{3}m+{\kfrac}n$& &$\frac{4}{3}m+5n$& & \\ 
\hline
  \begin{tabular}{c}
  degree-one free 
  \end{tabular}
  &&& &&$3m$& \\
\hline
  triconnected&&$2m+3n$& &$2m+3n$&& \\
\hline
  \begin{tabular}{c}
  simple \&\\triconnected
  \end{tabular}
  & &$2m+2n$& &$2m+2n$&$2.84m$&$\frac{3\log3}{2}m$\\
\hline
  triangulated& &$2m+2n$& &$2m+2n$&\Xomit{$1.53m$}& \\
\hline
  \begin{tabular}{c}
  simple \&\\triangulated
  \end{tabular}
  & &$2m+n$& &$2m+n$&$\frac{4}{3}m$&\\ 
\hline
\end{tabular}
\end{center}
\caption{This table compares our results with previous ones, where $n$
is the number of nodes, $m$ is the number of edges, and $k$ is a positive
constant. The lower-order terms are omitted.  All but row 1 assume
that $G$ has no self-loop.}
\end{figure*}

Our coding schemes employ two new tools. One is new techniques of
processing strings of multiple types of parentheses. This generalizes
the results on strings of a single type of parentheses in
\cite{MR97}. The other tool is new properties of canonical orderings
for plane graphs. Such orderings were introduced by de Fraysseix, Pach
and Pollack~\cite{DeFPP90} and extended by Kant~\cite{Kant92}. These
structures and closely related ones have proven useful also for
drawing plane graphs in organized and compact
manners~\cite{KH97,kaofhr94,Read87,Schnyder90}.

Section \ref{sec:tools} discusses the new tools. Section
\ref{sec:coding-with-query} describes the coding schemes that support
queries. Section \ref{compact_coding} presents the more compact coding
schemes which do not support queries.  The methods used in
\S\ref{sec:coding-with-query} and \S\ref{compact_coding} are
independent, and these two sections can be read in the reverse order.

{\it Remark.} Throughout this paper, for all our coding schemes, it is
straightforward to verify that both encoding and decoding take linear
time in the size of the input graph.  Hence for the sake of
conciseness, the corresponding theorems do not state this time
complexity.

\section{New Encoding Tools}\label{sec:tools}

\subsection{Basics}
A {\em simple} (respectively, {\em multiple}) graph is one that does
not contain (respectively, may contain) multiple edges between two
distinct nodes.  The {\em simple version} of a multiple graph is
one obtained from the multiple graph by deleting all but one copy of
each edge.

In this paper, all graphs are multiple and unlabeled unless explicitly
stated otherwise.  Furthermore, for technical simplicity, a multiple
graph is formally a simple one with positive integral edge weights,
where each edge's weight indicates its multiplicity.

The {\em degree} of a node $v$ in a graph is the number of edges,
counting multiple edges, incident to $v$ in the graph. A node $v$ is a
{\em leaf} of a tree $T$ if $v$ has exactly one neighbor in $T$. Since
$T$ may have multiple edges, a leaf of $T$ may have a degree greater
than one. We say $v$ is {\em internal} in $T$ if $v$ has more than one
neighbor in $T$.  See \cite{Ber85,GrotschelLS88} for other
graph-theoretic terminology used in this paper.

For a given problem of size $n$, this paper uses the $\log n$-bit word
model of computation in \cite{Jacobson89,Munro96,MR97}, where
operations such as read, write, add and multiply on $O(\log n)$
consecutive bits take $O(1)$ time.  The model can be implemented using
the following techniques.  If a given chunk of $O(\log n)$ consecutive
bits do not fit into a single word, they can be read or written by
$O(1)$ accesses of consecutive words.  Basic operations can be
implemented by table look-up methods similar to the Four Russian
algorithm~\cite{ahu74}.

\subsection{Multiple Types of Parentheses}\label{sec:multiple_strings}
A string is {\em binary} if it contains at most two kinds of symbols;
e.g., a string of one type of parentheses is a binary string.

\begin{fact}[see \cite{BCW90,Elias75}]\label{fact_aux}
Let $k=O(1)$.  Given any strings $S_1,S_2,\ldots,S_k$ with total
length $O(n)$, there exists an auxiliary binary string $\lambda$ such
that
\begin{itemize}
\item
the string $\lambda$ has $O(\log n)$ bits and can be computed in
$O(n)$ time;
\item 
given the concatenation of $\lambda,S_1,S_2,\ldots,S_k$ as input, the
index of the first symbol of any given $S_i$ in the concatenation can
be computed in $O(1)$ time.
\end{itemize}
\end{fact}
Let $S_1+S_2+\cdots+S_k$ denote the concatenation of $\lambda,
S_1,S_2,\ldots,S_k$ as in Fact~\ref{fact_aux}.

Let $S$ be a string. Let $|S|$ be the length of $S$.  Let $S[i]$ be
the symbol at the $i$-th position of $S$.  $S[k]$ is {\em enclosed} by
$S[i]$ and $S[j]$ in $S$ if $i<k<j$.  Let $\select{S,i,\Box}$ be the
position of the $i$-th $\Box$ in $S$. Let $\rank{S,k,\Box}$ be the
number of $\Box$'s before or at the $k$-th position of $S$. Clearly,
if $k=\select{S,i,\Box}$, then $i=\rank{S,k,\Box}$.

Now let $S$ be a string of multiple types of parentheses. For an open
parenthesis $S[i]$ and a close one $S[j]$ of the same type where
$i<j$, the two {\it match} in $S$ if every parenthesis of the same
type that is enclosed by them matches one enclosed by them.  $S$ is
{\em balanced} if every parenthesis in $S$ belongs to a matching
parenthesis pair.

Here are some queries defined for $S$:
\begin{itemize}
\item Let $\match{S,i}$ be the position of the parenthesis in $S$
that matches $S[i]$.

\item Let $\firstk{k}{S,i}$ (respectively, $\lastk{k}{S,i}$) be the
position of the first (respectively, last) parenthesis of the $k$-th
type after (respectively, before) $S[i]$.

\item Let $\enclosek{k}{S,i_1,i_2}$ be the positions $(j_1,j_2)$ of
the closest matching parenthesis pair of the $k$-th type that
encloses $S[i_1]$ and $S[i_2]$.
\end{itemize}
The answer to a query may be undefined; e.g., $\match{S,i}$ is
undefined for some $i$ if $S$ is not balanced.  If there is only one
type of parentheses in $S$, the subscript $k$ in $\firstk{k}{S,i}$,
$\lastk{k}{S,i}$, and $\enclosek{k}{S,i,j}$ may be omitted; thus,
$\firstk{}{S,i}=i+1$ and $\lastk{}{S,i}=i-1$. If it is clear from the
context, the parameter $S$ may also be omitted.

\begin{fact}[see \cite{Clark96,Munro96,MR97}]\
\label{fact:one-type}

\begin{enumerate}
\item\label{fact:one-type-one} Let $S$ be a binary string. An
auxiliary binary string $\mu_1(S)$ of length $o(|S|)$ is obtainable in
$O(|S|)$ time such that $\rank{S,i,\Box}$ and $\select{S,i,\Box}$ can
be answered from $S+\mu_1(S)$ in $O(1)$ time.
\item\label{fact:one-type-two} Let $S$ be a balanced string of one
type of parentheses. An auxiliary binary string $\mu_2(S)$ of length
$o(|S|)$ is obtainable in $O(|S|)$ time such that $\match{S,i}$ and
$\enclosek{}{S,i,j}$ can be answered from $S+\mu_2(S)$ in $O(1)$ time.
\end{enumerate}
\end{fact}

The next theorem generalizes Fact~\ref{fact:one-type}.

\begin{theorem}\label{theorem:more-types}
Let $S$ be a string of $O(1)$ types of parentheses that may be
unbalanced. An auxiliary $o(|S|)$-bit string $\aux(S)$ is obtainable
in $O(|S|)$ time such that $\rank{S,i,\Box}$, $\select{S,i,\Box}$,
$\firstk{k}{S,i}$, $\lastk{k}{S,i}$, $\match{S,i}$, and
$\enclosek{k}{S,i,j}$ can be answered from $S+\aux(S)$ in $O(1)$ time.
\end{theorem}

\begin{proof}
The case of $\rank{S,i,\Box}$ and $\select{S,i,\Box}$ is a
straightforward generalization of
Fact~\ref{fact:one-type}(\ref{fact:one-type-one}).  The case of
$\firstk{k}{S,i}$ is proved as follows.  Let $f(S,i,\Box)$ be the
position of the first $\Box$ after $S[i]$. Then,
\begin{eqnarray*}
f(S,i,\Box)&=&\select{S,1+\rank{S,i,\Box},\Box};\\
\firstk{k}{S,i}&=&\min\setof{f(S,i,\leftp{}),f(S,i,\rightp{})},
\end{eqnarray*}
where $\leftp{}$ and $\rightp{}$ are the open and close parentheses of
the $k$-th type in $S$, respectively. The case of $\lastk{k}{S,i}$ can
be shown similarly.

To prove the case of $\match{S,i}$ and $\enclosek{k}{S,i,j}$, we first
generalize Fact~\ref{fact:one-type}(\ref{fact:one-type-two}) for an
unbalanced binary $S$.  Let $R$ be the shortest balanced superstring
of $S$. Let $d=|R|-|S|$.  $R$ is either $S$ appended by $d$ close
parentheses or $d$ open parentheses appended by $S$.  Let $\beta(S)$
be $\mu_2(R)$ appended to $1+\lceil\log (n+1) \rceil$ bits which
record $d$ and whether $S$ is a prefix or a suffix of $R$. Then, a
query for $S$ can be answered from $S+\beta(S)$ in $O(1)$ time.

Now suppose that $S$ is of $\ell$ types of parentheses. Let $S_k$ with
$1\leq k\leq\ell$ be the string obtained from $S$ as follows.
\begin{itemize}
\item Every open (respectively, close) parenthesis of the $k$-th type
is replaced by two consecutive open (respectively, close) parentheses of
the $k$-th type. 
\item Every  parenthesis of any other type is replaced by a
matching parenthesis pair of the $k$-th type.
\end{itemize}
Each $S_k$ is a string of length $2|S|$ using one type of parentheses.
Each symbol $S_k[i]$ can be determined from $S[\floor{i/2}]$ in $O(1)$
time. For example,
\begin{center}
  \begin{tabular}{lcl}
     $S$  &=&\verb+[ [ ( { ) ] ( { } } ( ] ) +\\
     $S_1$&=&\verb+()()((()))()((()()()((()))+\\
     $S_2$&=&\verb+[[[[[][][]]][][][][][]]][]+\\
     $S_3$&=&\verb+{}{}{}{{{}{}{}{{}}}}{}{}{}+
  \end{tabular}
\end{center}
A query for $S$ can be answered by answering the queries for some
$S_k$ as follows.
\begin{itemize}
   \item $\match{S,i}=\floor{\match{S_k,2i}/2}$, where $S[i]$ is a
         parenthesis of the $k$-th type. 
   \item Let $i$ and $j$ be two positions. Let 
         \begin{eqnarray*}
            A &=&\setof{2i, 2i+1, \match{S_k,2i},\match{S_k,2i+1}} \cup \\
              & &\setof{2j, 2j+1, \match{S_k,2j},\match{S_k,2j+1}}.
         \end{eqnarray*}
         Let $i_1=\min A$,
         $j_1=\max A$, and $(i_2,j_2)=\enclose{S_k,i_1,j_1}$.
         Then, $\enclosek{k}{S,i,j}=(\floor{i_2/2},\floor{j_2/2})$.
\end{itemize}
Each query above on $S_k$ can be answered in $O(1)$ time from
$S_k+\beta(S_k)$. Since each symbol $S_k[i]$ can be determined from
$S[\floor{i/2}]$ in $O(1)$ time, the theorem holds by letting
$\aux(S)=\beta(S_1)+\beta(S_2)+\cdots+\beta(S_\ell)$.
\end{proof}

Given $k$ strings $S_1,\ldots,S_k$ of $O(1)$ types of parentheses, let
$\aux(S_1,S_2,\ldots,S_k)$ denote
$\aux(S_1)+\aux(S_2)+\cdots+\aux(S_k)$.

\subsection{Encoding Trees}
An encoding for a plane graph $G$ is {\em weakly convenient} if it
takes linear time to reconstruct $G$; $O(1)$ time to determine the
adjacency of two nodes in $G$; $O(d)$ time to determine the degree $d$
of a node; and $O(d)$ time to list the neighbors of a node of degree
$d$.  A weakly convenient encoding for $G$ is {\em convenient} if it
takes $O(1)$ time to determine the degree of a node.

For a simple rooted tree $T$, the folklore encoding $F(T)$ is defined
as follows. Initially, $F(T)$ is a balanced string of one type of
parentheses representing the preordering of $T$.  An open
(respectively, close) parenthesis denotes a descending (respectively,
ascending) edge traversal. Then, this string is enclosed by an
additional matching parenthesis pair. Note that each node of $T$
corresponds to a matching parenthesis pair in $F(T)$.

\begin{fact}\label{fact:string-property}
Let $v_i$ be the $i$-th node in the preordering of a simple rooted
tree $T$.
\begin{enumerate}
\item 
The parenthesis pair for $v_i$ encloses that for $v_j$ in $F(T)$ if
and only if $v_i$ is an ancestor of $v_j$.
\item\label{fact:string-property:2}
The parenthesis pair for $v_i$ precedes that for $v_j$ in $F(T)$ if
and only if $v_i$ and $v_j$ are not related and $i<j$.
\item\label{fact:string-property:3}
The $i$-th open parenthesis in $F(T)$ belongs to the parenthesis
pair for $v_i$.
\end{enumerate}
\end{fact}

\begin{fact}[see \cite{MR97}]\label{fact:folklore}
For a simple rooted tree $T$ of $n$ nodes, $F(T)+\mu_1(F(T))+\mu_2(F(T))$
is a weakly convenient encoding of $2n+o(n)$ bits.
\end{fact}

Based on Theorem~\ref{theorem:more-types}, we show that
Fact~\ref{fact:folklore} holds even if $F(T)$ is interleaved with
other types of parentheses.

\begin{theorem}\label{theorem:augmented-folklore} 
Let $T$ be a simple rooted tree.  Let $S$ be a string of $O(1)$ types
of parentheses such that a given type of parentheses in $S$ gives
$F(T)$. Then $S+\aux(S)$ is a weakly convenient encoding of $T$.
\end{theorem}
\begin{proof}
Let the parentheses, denoted by $\leftp{}$ and $\rightp{}$, in $S$
used by $F(T)$ be the $k$-th type. Let $v_1,\ldots,v_n$ be the
preordering of $T$.  Let $p_i=\select{S,i,\leftp{}}$ and
$q_i=\match{S,p_i}$; i.e., $S[p_i]$ and $S[q_i]$ are the matching
parenthesis pair corresponding to $v_i$ by Fact~\ref{fact:string-property}(\ref{fact:string-property:3}). By
Theorem~\ref{theorem:more-types}, each $p_i$ and $q_i$ are obtainable
from $S+\aux(S)$ in $O(1)$ time. Moreover, the index $i$ is obtainable
from $p_i$ or $q_i$ in $O(1)$ time by
$i=\rank{S,p_i,\leftp{}}=\rank{S,\match{S,q_i},\leftp{}}$.  The
queries for $T$ are supported as follows.

{\it Case} 1: adjacency queries. Suppose $i<j$. Then,
$(p_i,q_i)=\enclosek{k}{p_j,q_j}$ if and only if $v_i$ is adjacent to
$v_j$ in $T$, i.e., $v_i$ is the parent of $v_j$ in $T$.

{\it Case} 2: neighbor queries. Suppose that $v_i$ has degree $d$ in $T$. The
neighbors of $v_i$ in $T$ can be listed in $O(d)$ time as
follows. First, if $i\ne1$, output $v_j$, where
$(p_j,q_j)=\enclosek{k}{p_i,q_i}$.  Then, let
$p_j=\firstk{k}{p_i}$. As long as $p_j<q_i$, we repeatedly output
$v_j$ and update $p_j$ by $\firstk{k}{\match{p_j}}$.

{\it Case} 3: degree queries. Since $T$ is simple, the degree $d$ of
$v_i$ in $T$ is the number of neighbors in $T$, which is obtainable in
$O(d)$ time.
\end{proof}

The next theorem improves Theorem \ref{theorem:augmented-folklore} and
is important for our later coding schemes.  A related result in
\cite{MR97} shows that a $k$-page graph of $n$ nodes and $m$ edges has
a convenient encoding of $2m+2kn+o(m+n)$ bits. Since $T$ is a one-page
graph, this result gives a longer convenient encoding for $T$ than the
next theorem.

For a condition $P$, let $\delta(P)=1$, if $P$ holds; $\delta(P)=0$,
otherwise.

\begin{theorem}\label{theorem:degree}
Let $T$ be a rooted tree of $n$ nodes, $n^*$ leaves and $m$ edges. Let
$S+\aux(S)$ be a weakly convenient encoding of the simple version
$T_s$ of $T$.
\begin{enumerate} 
\item\label{thm:deg:1}
A string $D$ of $2m-n+n^*$ bits is obtainable in $O(m+n)$ time such
that $S+D+\aux(S,D)$ is a convenient encoding for $T$.
\item\label{thm:deg:2}
If $T=T_s$, a string $D$ of $n^*$ bits and a string $Y$ of $n$ bits
are obtainable in $O(m+n)$ time such that $S+D+\aux(S,D,Y)$ is a
convenient encoding for $T$.
\end{enumerate}
\end{theorem}
{\it Remark.}  In Statement~\ref{thm:deg:2}, the convenient encoding
contains $\aux(Y)$ but not $Y$ itself, which is only used in the
decoding process and is not explicitly stored. This technique is also
used in our other schemes.
\begin{proof}
Let $v_1,\ldots,v_n$ be the preordering of $T_s$. Let $d_i$ be the
degree of $v_i$ in $T$.  We show how to use $D$ to store the
information required to obtain $d_i$ in $O(1)$ time.

Statement 1.  Let $\delta_i=\delta(\mbox{$v_i$ is internal in
$T_s$})$. Since $S+\aux(S)$ is a weakly convenient encoding for $T_s$,
each $\delta_i$ is obtainable in $O(1)$ time from $S+\aux(S)$.
Initially, $D$ is $n$ copies of $1$. Let $b_i=d_i-1-\delta_i$.  We add
$b_i$ copies of $0$ right after the $i$-th 1 in $D$ for each
$v_i$. Since the number of internal nodes in $T_s$ is $n-n^*$, the bit
count of $D$ is $n+\sum_{i=1}^{n}(d_i-1-\delta_i)=2m-n+n^*$.  $D$ is
obtainable from $T$ in $O(m+n)$ time.  The number $b_i$ of 0's right
after the $i$-th 1 in $D$ is
$\select{D,i+1,\mbox{1}}-\select{D,i,\mbox{1}}-1$.  Since
$d_i=1+\delta_i+b_i$, the degree of $v_i$ in $T$ can be computed in
$O(1)$ time from $S+D+\aux(S,D)$.

Statement 2.  Let $n_2$ be the number of nodes of degree two in $T$.
Initially, $D$ is $n-n^*-n_2$ copies of $1$, one for each node of degree
at least three in $T$. Suppose that $v_i$ is the $h_i$-th node in
$v_1,\ldots,v_n$ of degree at least three. We put $d_i-3$ copies of
$0$ right after the $h_i$-th 1 in $D$. The bit count of $D$ is
$(n-n^*-n_2)+\sum_{i,d_i \geq 3} (d_i-3)= (n-n^*-n_2)+ (\sum_{i=1}^n
d_i - n^* -2n_2)-3(n-n^*-n_2)=n^*-2< n^*$.

Since $S+\aux(S)$ is a weakly convenient encoding for $T$, it takes
$O(1)$ time to determine whether $d_i\geq 3$ from $S+\aux(S)$.  If
$d_i<3$, $d_i$ can also be computed in $O(1)$ time from $S+\aux(S)$.
To compute $d_i$ when $d_i\geq 3$, note that since
$d_i=3+\select{D,h_i+1,\mbox{1}}-\select{D,h_i,\mbox{1}}-1$, it
suffices to compute $h_i$ in $O(1)$ time. Let $Y$ be an $n$-bit string
such that $Y[i]=1$ if and only if $d_i\geq3$. Then,
$h_i=\rank{Y,i,\mbox{1}}$, obtainable in $O(1)$ time from $Y+\aux(Y)$.
Each symbol $Y[i]$ can be determined from $S+\aux(S)$ in $O(1)$ time,
and we do not need to store $Y$ in our encoding.
\end{proof}

\subsection{Canonical Orderings}\label{section:canonical}
This section reviews {\em canonical orderings} of plane graphs
\cite{DeFPP90,Kant92} and proves new properties needed in our coding
schemes.

\begin{figure*}[t]
\centerline{\psfig{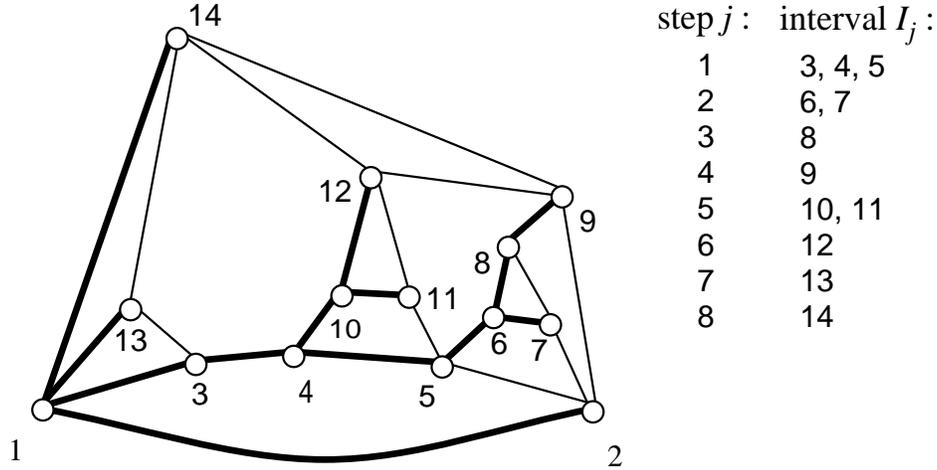}}
\caption{A triconnected plane graph $G$ and a canonical ordering of $G$.}
\label{canonical-triconnected}
\end{figure*}

\begin{figure}[t]
\centerline{\psfig{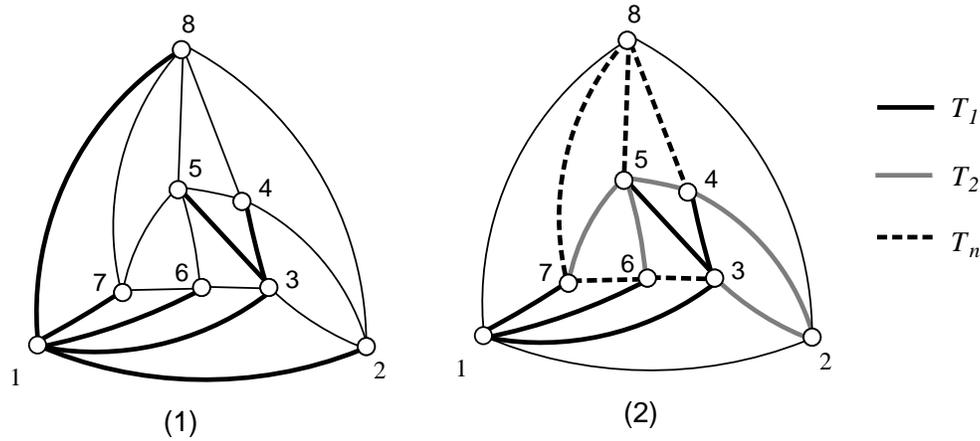}}
\caption{$(1)$ a canonical ordering of a plane triangulation $G$;
$(2)$ a realizer of $G$.}
\label{canonical-triangulate}
\end{figure}

All graphs in this section are simple.  Let $G$ be a plane graph. Let
$v_1,v_2,\ldots,v_n$ be an ordering of the nodes of $G$. Let $G_i$ be
the subgraph of $G$ induced by $v_1,v_2,\ldots,v_i$. Let $H_i$ be the
boundary of the exterior face of $G_i$.  This ordering is {\em
canonical} if the interval $[3,n]$ can be partitioned into
$I_1,\ldots,I_K$ with the following properties for each $I_j$.
Suppose $I_j=[k,k+q]$. Let $C_j$ be the path
$v_k,v_{k+1},\ldots,v_{k+q}$.
\begin{itemize}
\item 
$G_{k+q}$ is biconnected. $H_{k+q}$ contains the edge $(v_1,v_2)$ and
$C_j$. $C_j$ has no chord in $G$.  

{\it Remark.} Since $H_{k+q}$ is a cycle, to enhance visual
intuitions, we draw its nodes in the clockwise order from left to
right above the edge $(v_1,v_2)$.

\item 
If $q=0$, $v_k$ has at least two neighbors in $G_{k-1}$, all on
$H_{k-1}$.  If $q>0$, $C_j$ has exactly two neighbors in $G_{k-1}$,
both on $H_{k-1}$, where the left neighbor is incident to $C_j$ only
at $v_k$ and the right neighbor only at $v_{k+q}$.

{\it Remark.} Whether $q = 0$ or not, let $v_\ell$ and $v_r$ denote the
leftmost neighbor and the rightmost neighbor of $C_j$ on $H_{k-1}$.

\item 
For each $v_i$ where $k\leq i \leq k+q$, if $i < n$, $v_i$ has at
least one neighbor in $G-G_{k+q}$.
\end{itemize}
Figure~\ref{canonical-triconnected} shows a canonical ordering of a
triconnected plane graph; Figure~\ref{canonical-triangulate}(1)
illustrates one for a plane triangulation.
\begin{fact}[see \cite{DeFPP90,Kant92}]\
\label{fact_real}

\begin{enumerate}
\item \label{yyy1}
If $G$ is triconnected or triangulated, then it has a canonical
ordering that can be constructed in linear time.
\item 
For every canonical ordering of a triangulated $G$,
\begin{itemize}
\item  each $I_j$ consists of exactly one node, i.e., $q = 0$;
\item 
the neighbors of $v_k$ in $G_{k-1}$ form a subinterval of the path
$H_{k-1}-\{(v_1,v_2)\}$, where $H_2-\{(v_1,v_2)\}$ is regarded as the
edge $(v_1,v_2)$ itself.
\end{itemize}
\end{enumerate}
\end{fact}

Given a canonical ordering of $G$ with its unique partition
$I_1,I_2,\ldots,I_K$, $G=G_n$ is obtainable from $G_2 = \{(v_1,v_2)\}$
in $K$ steps, one for each $I_j$.  Step $j$ obtains $G_{k+q}$ from
$G_{k-1}$ by adding the path $v_k,v_{k+1},\ldots,v_{k+q}$ and its
incident edges to $G_{k-1}$. This process is called the {\em construction
algorithm} for $G$ corresponding to the ordering.

For the given ordering, the {\em canonical spanning tree} $T$ of $G$
rooted at $v_1$ is the one formed by the edge $(v_1,v_2)$ together
with the paths $C_j$ and the edges $(v_\ell,v_k)$ over all $I_j$.  In
Figures~\ref{canonical-triconnected}
and~\ref{canonical-triangulate}(1), $T$ is indicated by thick lines.
\begin{lemma}\
\label{lemma:canonical-tree}

\begin{enumerate}
\item \label{lemma:canonical-tree:statement1}
For every edge $(v_i,v_{i'})$ in $G-T$, $v_i$ and $v_{i'}$ are not related in
$T$.
\item \label{lemma:canonical-tree:statement2}
For each node $v_i$, the edges incident to $v_i$ in $G$ form the
following pattern around $v_i$ in the counterclockwise order: an edge
to its parent in $T$; followed by a block of nontree edges to
lower-numbered nodes; followed by a block of tree edges to its
children in $T$; followed by a block of nontree edges to
higher-numbered nodes, where a block may be empty.
\end{enumerate}
\end{lemma}
\begin{proof}

Statement 1. Suppose that $(v_i,v_{i'})$ is added at step $j$ of the
construction algorithm for $G$.  Either $v_i$ or $v_{i'}$ is on the
path $v_k,v_{k+1},\ldots,v_{k+q}$ and the other is to the right of
$v_\ell$ on $H_{k-1}$. Hence $v_i$ is neither an ancestor nor a
descendant of $v_{i'}$ in $T$.

Statement 2. Suppose that $v_i$ is added at step $j$.  The tree edge
from $v_i$ to its parent, i.e., $v_\ell$ or $v_{i-1}$, and all the
nontree edges between $v_i$ and its lower-numbered neighbors are added
during this step; if $i < k+q$, no such nontree edge exists.  All such
nontree edges precede other edges incident to $v_i$ in the
counterclockwise order.  Any edge $e=(v_i,v_{i'})$ with $i < i'$ is
added during step $j'$ with $j< j'$.  Let $I_{j'}=[k',k'+q']$. 
Thus, $e$ is a tree edge only if
$i'=k'$ and $v_i$ is the leftmost neighbor of $v_{i'}$ in $H_{i'-1}$
for otherwise $e$ would be a nontree edge.  The tree edges between
$v_i$ and its children, which are higher numbered, precede the nontree
edges between $v_i$ and its higher-numbered neighbors in the
counterclockwise order.
\end{proof}

Let $T'$ be a tree embedded on the plane. Let $(x,y)$ be an edge of
$T'$.  The {\em counterclockwise preordering of $T'$ starting at $x$ and
$y$} is defined as follows. We perform a preorder traversal on $T'$
starting at $x$ and using $(x,y)$ as the first visited edge.  Once a
node $v$ is visited via $(w,v)$, the unvisited nodes adjacent to $v$ are
visited in the counterclockwise order around $v$ starting from the first
edge following $(w,v)$.

\begin{fact}[see \cite{Kant93,hkaol.coding_one}] 
\label{fact:preordering} 
For every triconnected plane graph, the counterclockwise preordering
of any canonical spanning tree is also a canonical ordering of the
graph.
\end{fact}

{\it Remark.} The canonical ordering in
Figure~\ref{canonical-triconnected} is the counterclockwise
preordering of $T$.

Assume that $G$ is a triangulation with exterior nodes $v_1,v_2,v_n$
in the counterclockwise order. A {\em realizer} of $G$ is a partition
of the interior edges of $G$ into three trees $T_1,T_2,T_n$ rooted at
$v_1,v_2,v_n$ respectively with the following properties
\cite{Schnyder90}:
\begin{enumerate}
\item 
All the interior edges incident to $v_1$ ($v_2$ or $v_n$,
respectively) belong to $T_1$ ($T_2$ or $T_n$, respectively) and oriented
to $v_1$ ($v_2$ or $v_n$, respectively).
\item For each interior node $v$, the edges incident to $v$ form the
following pattern around $v$ in the counterclockwise order:
an edge in $T_1$ leaving $v$; followed by
       a block of edges in $T_n$ entering $v$;
an edge in $T_2$ leaving $v$; followed by
       a block of edges in $T_1$ entering $v$;
an edge in $T_n$ leaving $v$; followed by
       a block of edges in $T_2$ entering $v$,
where a block may be empty.
\end{enumerate}
Figure~\ref{canonical-triangulate}(2) illustrates a realizer of the
plane triangulation of Figure~\ref{canonical-triangulate}(1).  The
next fact relates a canonical ordering and a realizer via
counterclockwise tree preordering.

\begin{fact}[see  \cite{Kant93,Schnyder90}] \label{fact:realizer}
Let $G$ be a plane triangulation.
\begin{enumerate}
\item \label{xxx1}
Let $v_1,v_2,\ldots,v_n$ be a canonical ordering of $G$.  Note that
each $I_j$ consists of a node $v_k$.  Orient and partition the
interior edges of $G$ into three subsets $T_1,T_2,T_n$ as follows. For
each $v_k$ with $ k \geq 3$, $(v_k,v_\ell)$ is in $T_1$ oriented to
$v_\ell$; $(v_k,v_r)$ is in $T_2$ oriented to $v_r$; the edges
$(v_k,v_i)$ where $\ell < i < r$ are in $T_n$ oriented to $v_k$.  Then
$T_1,T_2,T_n$ is a realizer of $G$.  Consequently, every plane
triangulation has a realizer that can be constructed in linear time.

\item \label{xxx2}
For a realizer $T_1,T_2,T_n$ of $G$, let $T=T_1 \cup
\{(v_1,v_2),(v_1,v_n)\}$.  Let $v_1,v_2,\ldots,v_n$ be the
counterclockwise preordering of $T$ that starts at $v_1$ and uses
$(v_1,v_2)$ as the first visited edge. Then $v_1,v_2,\ldots,v_n$ is a
canonical ordering of $G$, and $T$ is a canonical spanning tree
rooted at $v_1$.
\end{enumerate}
\end{fact}

In Figure~\ref{canonical-triangulate}(1), the tree $T$ stated in
Fact~\ref{fact:realizer}(\ref{xxx2}) is indicated by thick lines, and the
canonical ordering shown is the counterclockwise preordering of $T$.

\section{Schemes with Query Support}\label{sec:coding-with-query}
This section presents our coding schemes that support queries.  We
give a weakly convenient encoding 
in~\S\ref{subsection:basis}. This encoding illustrates the
basic techniques applicable to our coding schemes with query
support. We then give the schemes for triconnected, triangulated, and
general plane graphs in~\S\ref{subsection:triconnected-with-query},
~\S\ref{subsection:triangulated-with-query},
and~\S\ref{subsection:plane-with-query}, respectively. We show how to
accommodate self-loops in~\S\ref{sec:self-loops}.

\subsection{Basic Techniques}\label{subsection:basis}
\begin{itemize}
\item 
Let $G_s$ be a simple plane graph with $n$ nodes and $m_s$ edges. 
\item 
Let $T$ be a spanning tree of $G_s$ that satisfies
Lemma~\ref{lemma:canonical-tree}. Let $n^*$ be the number of leaves in
$T$. Let $v_1,\ldots,v_n$ be the counterclockwise preordering of $T$.
\item 
Let $G_a$ be a graph obtained from $G_s$ by adding multiple edges
between adjacent nodes in $G_s-T$. Let $m_a$ be the number of edges in
$G_a$, counting multiple edges.
\end{itemize}
We now give a weakly convenient encoding for $G_a$ using {\em parentheses} to
encode $T$ and {\em brackets} to encode the edges in $G_a-T$.  Initially, let
$S=F(T)$. Let $\leftp{i}$ and $\rightp{i}$ be the parenthesis pair
corresponding to $v_i$ in $S$. We insert into $S$ a pair $\leftb{e}$
and $\rightb{e}$ for each edge $e=(v_i,v_j)$ of $G_a-T$ with $i<j$ as
follows.
\begin{itemize}
   \item $\leftb{e}$ is placed right after $\rightp{i}$, and
   \item $\rightb{e}$ is placed right after $\leftp{j}$.
\end{itemize}
For example, the string $S$ for the graph in
Figure~\ref{canonical-triangulate} is:
\begin{center}
\verb+(()[[[(](])[[(])[[[)[(]])[(]])[(]]]]))+\\
\verb+122   3 4 4  5 5   3 6  6 7  7 8    81+
\end{center}
Note that if $v_i$ is adjacent to $\ell_i$ lower-numbered nodes and
$h_i$ higher-numbered nodes in $G_a-T$, then in $S$ 
the open parenthesis $\leftp{i}$ is immediately followed by
$\ell_i$ close brackets, and the close parenthesis $\rightp{i}$ by
$h_i$ open brackets.
\begin{lemma}\label{property_A}
The last parenthesis that precedes an open $($respectively, close$)$
bracket in $S$ is close $($respectively, open$)$.
\end{lemma}
\begin{proof}
Straightforward.
\end{proof}

\newcommand{\ehh}[0]{h}

\begin{figure}
\centerline{\psfig{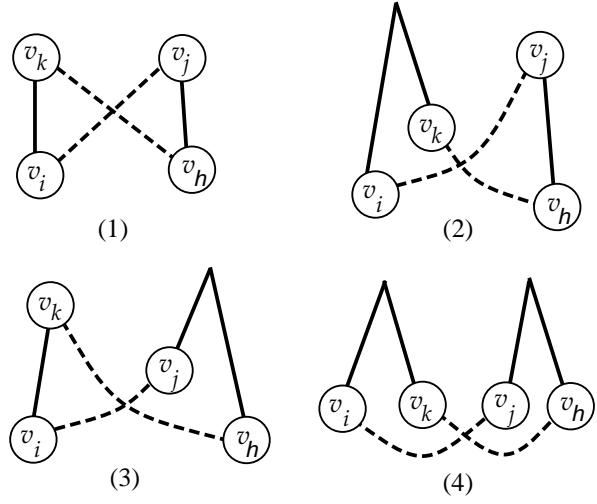}}
\caption{Edge crossing}
\label{figure:crossing}
\end{figure}

Let $e=(v_i,v_j)$ be an edge of $G_a-T$ with $i<j$. By
Lemma~\ref{lemma:canonical-tree}(\ref{lemma:canonical-tree:statement1}),
$v_i$ and $v_j$ are not related. By 
Fact~\ref{fact:string-property}(\ref{fact:string-property:2}),
$\rightp{i}$ precedes $\leftp{j}$ in $S$. Also, $\leftb{e}$ precedes
$\rightb{e}$ in $S$ for every edge $e$ in $G_a-T$, counting multiple
edges.  Note that $\leftb{e}$ and $\rightb{e}$ do not necessarily
match each other in $S$. In the next lemma, let $S[p]<S[q]$ denote
that $S[p]$ precedes $S[q]$ in $S$, i.e., $p<q$.
\begin{lemma}\label{lemma:xxx}
Let $e$ and $f$ be two edges in $G_a-T$ with no common endpoint.  If
$\leftb{e}<\leftb{f}$, then either
$\leftb{e}<\rightb{e}<\leftb{f}<\rightb{f}$ or
$\leftb{e}<\leftb{f}<\rightb{f}<\rightb{e}$.
\end{lemma}
\begin{proof}
Suppose $e=(v_i,v_j)$ and $f=(v_k,v_{\ehh})$, where $i<j$ and $k<{\ehh}$.
Assume for a contradiction that
$\leftb{e}<\leftb{f}<\rightb{e}<\rightb{f}$.  Since $e$ and $f$ have
no common endpoint, $\rightp{i}<\rightp{k}<\leftp{j}<\leftp{{\ehh}}$.
There are four possible cases:
\begin{enumerate}
\item
$\leftp{k}<\leftp{i}<\rightp{i}<\rightp{k}<\leftp{j}<\leftp{{\ehh}}<\rightp{{\ehh}}<\rightp{j}$; 
see Figure~\ref{figure:crossing}(1).
\item
$\leftp{i}<\rightp{i}<\leftp{k}<\rightp{k}<\leftp{j}<\leftp{{\ehh}}<\rightp{{\ehh}}<\rightp{j}$;
see Figure~\ref{figure:crossing}(2).
\item
$\leftp{k}<\leftp{i}<\rightp{i}<\rightp{k}<\leftp{j}<\rightp{j}<\leftp{{\ehh}}<\rightp{{\ehh}}$;
see Figure~\ref{figure:crossing}(3).
\item
$\leftp{i}<\rightp{i}<\leftp{k}<\rightp{k}<\leftp{j}<\rightp{j}<\leftp{{\ehh}}<\rightp{{\ehh}}$;
see Figure~\ref{figure:crossing}(4).
\end{enumerate}
In Figure~\ref{figure:crossing}, the dark lines are paths in $T$ and
the dashed ones are edges in $G_a-T$. The relation among these lines
follows from Fact \ref{fact:string-property} and
Lemma~\ref{lemma:canonical-tree}(\ref{lemma:canonical-tree:statement2}).
In all the cases, $e$ crosses $f$,
contradicting the fact that $G_a$ is a plane graph.
\end{proof}

By Lemma~\ref{lemma:xxx}, $\rightb{e}$ and the bracket that matches
$\leftb{e}$ in $S$ are in the same block of brackets.  From here
onwards, we rename the close brackets by redefining $\rightb{e}$ to be
the close bracket that matches $\leftb{e}$ in $S$. Note that
Lemma~\ref{property_A} still holds for $S$.

\begin{lemma}
\label{lemma:convenient-simple-triconnected}
$S+\aux(S)$ is a weakly convenient encoding for $G_a$. 
\end{lemma}
\begin{proof}
Since $T$ is simple, by Theorem~\ref{theorem:augmented-folklore}
$S+\aux(S)$ is a weakly convenient encoding for $T$. We next show that
$S+\aux(S)$ is also a weakly convenient encoding for $G_a-T$. Let
$p_i$ and $q_i$ be the positions of $\leftp{i}$ and $\rightp{i}$ in
$S$, respectively.

{\it Case} 1: adjacency queries. Suppose $i<j$. Then, $v_i$ and $v_j$
are adjacent in $G_a-T$ if and only if $q_i<p<q<\firstk{1}{p_j}$, where
$(p,q)=\enclosek{2}{\firstk{1}{q_i},p_j}$ as shown below.
\begin{displaymath}
\begin{array}{cccccc} 
   \rightp{i} & \leftb{} &  & \leftp{j} & \rightb{}\\
   \uparrow & \uparrow & \uparrow & \uparrow & \uparrow & \uparrow\\
   q_i & p & \firstk{1}{q_i} & p_j & q & \firstk{1}{p_j}
\end{array}
\end{displaymath}

{\it Case} 2: neighbor and degree queries. The neighbors and thus the
degree of a degree-$d$ node $v_i$ in $G_a-T$ are obtainable in $O(d)$
time as follows.

For each position $p$ such that $q_i < p < \firstk{1}{q_i}$, we
output $v_j$, where $p_j=\lastk{1}{\match{p}}$ as shown below.  Note
that $(v_i,v_j)$ is an edge in $G_a-T$ with $j > i$.
\begin{displaymath}
\begin{array}{ccccc} 
   \rightp{i} & \leftb{} &  & \leftp{j} & \rightb{}\\
   \uparrow & \uparrow & \uparrow & \uparrow & \uparrow\\
   q_i & p & \firstk{1}{q_i} & p_j & \match{p}
\end{array}
\end{displaymath}

For each position $q$ such that $p_i < q < \firstk{1}{p_i}$, we
output $v_j$, where $q_j=\lastk{1}{\match{q}}$ as shown below.  Note
that $(v_i,v_j)$ is an edge in $G_a-T$ with $j < i$.
\begin{displaymath}
\begin{array}{cccccc} 
     \rightp{j} &\leftb{}   & \leftp{i} &\rightb{}\\
     \uparrow   & \uparrow  & \uparrow  & \uparrow  & \uparrow\\
     q_j        &\match{q}& p_i       & q         & \firstk{1}{p_i}
\end{array}
\end{displaymath}
\end{proof}

Since $|S|=2n+2(m_a-n+1)=2m_a+2$ and $S$ uses four symbols, $S$ can be
encoded by $4m_a+4$ bits. The next lemma improves this bit count.
\begin{lemma}
\label{lemma:better-bit-count}
Let $S'$ be a string of $s_1$ parentheses and $s_2$ brackets that
satisfies Lemma~\ref{property_A}. Then $S'$ can be encoded by a string
of $2s_1+s_2+o(s_1+s_2)$ bits, from which each $S'[i]$ can be
determined in $O(1)$ time.
\end{lemma}
\begin{proof}
Let $S'_1$ and $S'_2$ be two binary strings
defined as follows, both obtainable in $O(|S'|)$ time:
\begin{itemize}
\item $S'_1[i]=1$ if and only if $S'[i]$ is a parenthesis for $1\leq
i\leq s_1+s_2$;
\item $S'_2[j]=1$ if and only if the $j$-th parenthesis in $S'$ is open
for $1\leq j\leq s_1$.
\end{itemize}
Each $S'[i]$ can be determined from $S'_1+S'_2+\aux(S'_1)$ in $O(1)$ time
as follows. Let $j=\rank{S'_1,i,\mbox{1}}$. If $S'_1[i]=1$, $S'[i]$ is a
parenthesis. Whether it is open or close can be determined from
$S'_2[j]$. If $S'_1[i]=0$, $S'[i]$ is a bracket. Whether it is open or
close can be determined from
$S'_2[\select{S'_1,\rank{S'_1,i,\mbox{1}},\mbox{1}}]$ by
Lemma~\ref{property_A}.
\end{proof}

The next lemma summarizes the above discussion.
\begin{lemma}
\label{lemma:weakly-convenient-simple-triconnected}
$G_a$ has a weakly convenient encoding of $2m_a+2n+o(m_a+n)$ bits,
from which the degree of a node in $G_a-T$ is obtainable in $O(1)$
time.
\end{lemma}
\begin{proof}
This lemma follows from
Lemmas~\ref{property_A},~\ref{lemma:convenient-simple-triconnected},
and~\ref{lemma:better-bit-count} and the fact that $S$ contains $2n$
parentheses and $2(m_a-n+1)$ brackets. 
\end{proof}

\subsection{Triconnected Plane Graphs}
\label{subsection:triconnected-with-query}
This section adopts all the notation of~\S\ref{subsection:basis} with
the following further definitions.
\begin{itemize} 
\item 
Let $G$ be a triconnected plane graph.
\item 
Let $G_s$ be the simple version of $G$.
\item 
Let $T$ be a canonical spanning tree of $G_s$, which therefore
satisfies Lemma~\ref{lemma:canonical-tree}.
\end{itemize}
Note that $G$ is obtained from $G_a$ be adding multiple edges between 
adjacent nodes in $T$.
We next show that the weakly convenient encoding for $G_a$ in
Lemma~\ref{lemma:weakly-convenient-simple-triconnected} can be
shortened to $2(m_a+n-n^*)+o(n)$ bits. We also give a convenient
encoding for $G_a$ of $2m_a+2n+o(n)$ bits. Then we augment both
encodings to accommodate multiple edges in $T$. This gives encodings
of $G$.

Let $v_h$ be a leaf of $T$ with $2<h<n$.  By the definitions of $T$
and a canonical ordering, $v_h$ is adjacent to at least one
higher-numbered node and at least two distinct lower-numbered nodes in
$G_a$. By the definition of $T$, the parent of $v_h$ in $T$, i.e., the
only neighbor of $v_h$ in $T$, has a lower number than $v_h$. Thus,
$v_h$ is adjacent to a higher-numbered node and a lower-numbered one
in $G_a-T$. Thus, $\leftp{h}$ is immediately succeeded by a
$\rightb{}$, and $\rightp{h}$ by a $\leftb{}$.
With these observations, we can remove a pair of brackets for every
$v_h$ from $S$ without losing any information on $G_a$ as
follows.  Let $P$ be the string obtained from $S$ by removing the
$\rightb{}$ that immediately succeeds $\leftp{h}$ as well as the
$\leftb{}$ that immediately succeeds $\rightp{h}$ for every 
$v_h$.  Let $Q$ be the string obtained from $S$ by removing
$\leftp{h}$ and $\rightp{h}$ for every $v_h$. 

Note that $|P|=|Q|$. Also, $Q[i]$ is obtainable from $P+\aux(P)$ in
$O(1)$ time as follows:
\begin{displaymath}
Q[i]=\left\{
  \begin{array}{ll}
     \rightb{}&\mbox{if $P[i]=\leftp{}$, $P[\firstk{1}{P,i}]=\rightp{}$, and $2<\rank{P,i,\leftp{}}<n$};\\
     \leftb{}&\mbox{if $P[i]=\rightp{}$, $P[\lastk{1}{P,i}]=\leftp{}$, and $2<\rank{P,i,\leftp{}}<n$};\\
     P[i]&\mbox{otherwise}.
  \end{array}
\right.
\end{displaymath}

\begin{lemma}
\label{lemma:p-and-q}
$P+Q+\aux(P,Q)$ is a weakly convenient encoding for $G_a$, and a
convenient encoding for $G_a-T$. 
\end{lemma}
\begin{proof}
Note that the parentheses in $P$ form $F(T)$. Thus, by
Theorem~\ref{theorem:augmented-folklore}, it suffices to show that
$P+Q+\aux(P,Q)$ is a convenient encoding for $G_a-T$ as follows. 

{\it Case} 1: adjacency queries. Given $i<j$, let
$(p,q)=\enclosek{2}{Q,\firstk{1}{P,q_i},p_j-1}$; the $-1$ in the last
parameter accounts for the possibility that $Q[p_j]$ is a bracket.
Note that $v_i$ is adjacent to $v_j$ if and only if $q_i\leq
p<q<\firstk{1}{P,p_j}$ as shown below. Here, the first inequality
accounts for the possibility of $Q[q_i]$ being a bracket.
\begin{displaymath}
\begin{array}{ccccccc} 
  P& \rightp{i} &          &                   & \leftp{j}\\
  Q&            & \leftb{} &                   &          & \rightb{}\\
   & \uparrow   & \uparrow & \uparrow          & \uparrow & \uparrow & \uparrow\\
   & q_i        & p        & \firstk{1}{P,q_i} & p_j      & q & \firstk{1}{P,p_j}
\end{array}
\end{displaymath}

{\it Case} 2: neighbor queries. The neighbors of $v_i$ can be listed as
follows.

For every position $p$ with
$q_i-\delta(Q[q_i]=\leftb{})<p<\firstk{1}{P,q_i}$, we output $v_j$,
where $p_j=\lastk{1}{P,\match{Q,p}+1}$ as shown below. Note that the
$+1$ in the last parameter accounts for the possibility of
$P[\match{Q,p}]$ being a parenthesis.  Also, $(v_i,v_j)$ is an edge in
$G_a-T$ with $j>i$.
\begin{displaymath}
\begin{array}{ccccccc} 
   P & \rightp{i} &          &  & \leftp{j} \\
   Q &            & \leftb{} &  &           & \rightb{}\\
     & \uparrow & \uparrow & \uparrow & \uparrow & \uparrow\\
     & q_i & p & \firstk{1}{P,q_i} & p_j & \match{Q,p}
\end{array}
\end{displaymath}

For every position $q$ with
$p_i-\delta(Q[p_i]=\rightb{})<q<\firstk{1}{P,p_i}$, we output $v_j$,
where $q_j=\lastk{1}{P,\match{Q,q}+1}$ as shown below. Note that the
$+1$ in the last parameter accounts for the possibility of
$P[\match{Q,q}]$ being a parenthesis.  Note that $(v_i,v_j)$ is an
edge in $G_a-T$ with $j<i$.
\begin{displaymath}
\begin{array}{ccccccc} 
   P & \rightp{j} &           & \leftp{i} &           &          \\
   Q &            &\leftb{}   &          & \rightb{} &          \\
     & \uparrow   & \uparrow  & \uparrow  & \uparrow  & \uparrow\\
     & q_j        &\match{Q,q}& p_i       & q         & \firstk{1}{P,p_i}
\end{array}
\end{displaymath}

{\it Case} 3: degree queries. The degree of $v_i$ in $G_a-T$ is
$\firstk{1}{P,q_i}-q_i+\delta(Q[q_i]=\leftb{})+\firstk{1}{P,p_i}-p_i+\delta(Q[p_i]=\rightb{})-2$,
obtainable from $P+Q+\aux(P,Q)$ in $O(1)$ time.
\end{proof}

\begin{lemma}\label{lemma:yyy}
$G_a$ has a weakly convenient encoding of $2m_a+2n-2n^*+o(m_a+n)$
bits, from which the degree of a node in $G_a-T$ is obtainable in
$O(1)$ time.  Moreover, $G_a$ has a convenient encoding of
$2m_a+2n-n^*+o(m_a+n)$ bits.
\end{lemma}
\begin{proof}
Since each $Q[i]$ is obtainable from $P+\aux(P)$ in $O(1)$ time, by
Lemma~\ref{lemma:p-and-q}, $P+\aux(P,Q)$ is also a weakly convenient
encoding for $G_a$.  Since $S$ satisfies Lemma~\ref{property_A} and
$P$ is obtained from $S$ by removing some brackets, $P$ also satisfies
Lemma~\ref{property_A}. Since $P$ has $2n$ parentheses and
$2(m_a-(n-1)-n^*)$ brackets, by Lemma~\ref{lemma:better-bit-count},
$G_a$ has a weakly convenient encoding of $2(m_a+n-n^*)+o(m_a+n)$
bits.
To augment this weakly convenient encoding into a convenient one, note
that the degree of $v_i$ in $G_a-T$ is obtainable in $O(1)$ time from
$P+Q+\aux(P,Q)$. By Theorem~\ref{theorem:degree}(\ref{thm:deg:2}), $n^*+o(n)$
additional bits suffice for supporting a degree query for $T$ in
$O(1)$ time.  Thus, $G_a$ has a convenient encoding of
$2m_a+2n-n^*+o(m_a+n)$ bits.
\end{proof}

The next theorem summarizes the above discussion and extends
Lemma~\ref{lemma:yyy} to accommodate multiple edges in $T$.
\begin{theorem}
\label{theorem:convenient-triconnected}
Let $G$ be a triconnected plane graph of $n$ nodes and $m$ edges.  Let
$G_s$ be the simple version of $G$ with $m_s$ edges.  Let $n^*$ be the
number of leaves in a canonical spanning tree $T$ of $G_s$.  Then $G$
$($respectively, $G_s$$)$ has a convenient encoding of
$2m+3n-n^*+o(m+n)$ $($respectively, $2m_s+2n-n^*+o(n)$$)$ bits.
\end{theorem}
\begin{proof}
The statement for $G_s$ follows immediately from Lemma~\ref{lemma:yyy}
with $G_a=G_s$.

To prove the statement for $G$, let $G_a$ be the graph obtained from $G_s$
by adding the multiple edges of $G$ between adjacent nodes in $G_s-T$.
By Lemma~\ref{lemma:yyy}, if
$G_a$ has $m_a$ edges, then $G_a$ has a weakly convenient encoding of
$2(m_a+n-n^*)+o(m_a+n)$ bits, from which a degree query for $G_a-T$
takes $O(1)$ time. 
Next, let $T_b=G-(G_a-T)$.  To support degree queries for $T_b$, note
that $T_b$ is a multiple tree of $n$ nodes and $m-m_a+n-1$ edges.  By
Theorem~\ref{theorem:degree}(\ref{thm:deg:1}), $2(m-m_a+n-1)-n+n^*+o(m)$
additional bits 
suffice for supporting a degree query of $T_b$ in $O(1)$ time.  Thus,
$G$ has a convenient encoding of $2m+3n-n^*+o(m)$ bits.
\end{proof}

\subsection{Plane Triangulations}
\label{subsection:triangulated-with-query}
Since every plane triangulation is triconnected, all the coding
schemes of Theorem~\ref{theorem:convenient-triconnected} are
applicable to plane triangulations. The next theorem shortens their
encodings.  The theorem and its proof adopt the notation of
\S\ref{subsection:basis} and
\S\ref{subsection:triconnected-with-query}.

\begin{theorem}
\label{theorem:triangulated}
Assume that $G$ is a plane triangulation of $n$ nodes and $m$ edges.
Let $G_s$ be the simple version of $G$ with $m_s=3n-6$ edges. Then $G$
$($respectively, $G_s$$)$ has a convenient encoding of $2m+2n+o(m+n)$
$($respectively, $2m_s+n+o(n)$$)$ bits.
\end{theorem}
\begin{proof}
By the definition of a canonical ordering, every $v_i$ with $1<i<n$ is
adjacent to a higher-numbered and a lower-numbered node in
$G_s-T$. Thus when computing the $P$ of
\S\ref{subsection:triconnected-with-query} from $S$, we can also
remove the $\leftb{}$ right after $\rightp{i}$ even if $v_i$ is
internal in $T$. Then, the string $Q$ of length $|P|$ is redefined as
follows:
\begin{displaymath}
Q[i]=\left\{
  \begin{array}{ll}

     \rightb{}&\mbox{if $P[i]=\leftp{}$,
     $P[\firstk{1}{P,i}]=\rightp{}$, and $2<\rank{P,i,\leftp{}}<n$};\\
     \leftb{}&\mbox{if $P[i]=\rightp{}$ and $1<\rank{P,i,\leftp{}}<n$};\\
     P[i]&\mbox{otherwise}.
  \end{array}
\right.
\end{displaymath}
The proof of Lemma~\ref{lemma:p-and-q} works identically. Since the
count of brackets decreases by $n-n^*+O(1)$, each encoding in
Theorem~\ref{theorem:convenient-triconnected} has $n-n^*+O(1)$ fewer
bits.
\end{proof}

\subsection{General Plane Graphs}\label{subsection:plane-with-query}
This section assumes that if a plane graph has more than one connected
component, then no connected component is inside an interior face of
another connected component.

Let $\hat{G}_s$ be a simple plane graph with $n$ nodes, $\hat{m}_s$
edges, and $c$ connected components
$\hat{M}_1,\hat{M}_2,\ldots,\hat{M}_c$.  Let $\hat{n}_j$ and
$\hat{m}_j$ be the numbers of nodes and edges in $\hat{M}_j$.

For each $\hat{M}_j$, we define a graph $M_j$ as follows.  If
$\hat{n}_j<3$, let $M_j = \hat{M}_j$. If $\hat{n}_j\geq 3$, let $M_j$
be a graph obtained by triangulating $\hat{M}_j$. Among the
$3\hat{n}_j-6$ edges in $M_j$, the ones in $\hat{M}_j$ are called {\em
real}, and the others are {\em unreal}.

For each $M_j$, we define a spanning tree $T_j$ as follows.  If
$\hat{n}_j<3$, let $T_j$ be an arbitrary rooted spanning tree of
$M_j$. For $\hat{n}_j\geq 3$, recall that by
Fact~\ref{fact:realizer}(\ref{xxx1}), $M_j$ has a realizer formed by
three edge-disjoint trees. Furthermore, three canonical spanning trees
$T_j^1, T_j^2, T_j^3$ of $M_j$ are obtainable by adding to each of these
three trees two boundary edges of the exterior face of $M_j$. Let
$T_j$ be a tree among $T_j^1, T_j^2, T_j^3$ with the least number of
unreal edges.

Let $T$ be the tree rooted at a new node $v_0$ by joining the root of
each $T_j$ to $v_0$ with an {\it unreal} edge; note that $T$ is
obtainable in $O(n)$ time by Fact~\ref{fact:realizer}.
Let $m_u$ be the number of the unreal edges of $T$; thus, $T$ has
$n-m_u$ real edges.  Let $v_0,v_1,v_2,\ldots,v_n$ be a
counterclockwise preordering of $T$.  Let $d_i$ be the degree of $v_i$
in $T$.  Let $N_k$ be the number of nodes of degree more than $k$ in
$T$.

Let $G_s$ be the simple graph composed of the edges in $\hat{G}_s$ and
the unreal edges in $T$.  A node of $G_s$ is {\em real} if its
incidental edge to its parent in $T$ is real; note that each child of
$v_0$ in $T$ is unreal.  

Let $E_a$ be a set of $\ell_a$ multiple edges between adjacent nodes
in $G_s-T$. Let $G_a=G_s\cup E_a$ and $\hat{G}_a=\hat{G}_s\cup E_a$.
Let $m_a$ and $\hat{m}_a$ be the numbers of edges in $G_a$ and
$\hat{G}_a$, respectively; i.e., $m_a=m_s+\ell_a$ and
$\hat{m}_a=\hat{m}_s+\ell_a$.

\begin{lemma}\ 
\label{eq:mu-1}

\begin{enumerate}
\item    \label{rel1} $m_u \leq n-\frac{1}{3}\hat{m}_s.$
\item    \label{rel2} $m_u-c \leq \frac{2}{3}n.$
\item    \label{rel3} $N_k \leq \frac{n}{k}.$
\end{enumerate}
\end{lemma}
\begin{proof}

Statement \ref{rel1}.  Let $u_j$ be the number of unreal edges of
$T_j$.  Clearly $m_u=c+u_1+u_2+\cdots+u_c$. Since
$\hat{m}_s=\hat{m}_1+\hat{m}_2+\cdots+\hat{m}_c$, it suffices to prove
the claim that $u_j\leq \hat{n}_j-\frac{1}{3}\hat{m}_j-1$ for every $j
= 1, 2, \ldots, c$.  For $\hat{n}_j\leq 2$, the claim holds
trivially. Now suppose $\hat{n}_j\geq 3$. For $t=1,2,3$, let $r_t$ and $u_t$
be the numbers of real and unreal edges in $T_j^t$, respectively. Since the
three trees in a realizer of $M_j$ are edge disjoint,
$r_1+r_2+r_3+u_1+u_2+u_3-6=3\hat{n}_j-9$. Since
$r_1+r_2+r_3\geq\hat{m}_j$ and $u_1+u_2+u_3\geq 3u_j$, the claim
holds.

Statement \ref{rel2}.
\begin{displaymath}
   m_u-c
   =\sum_{1\leq j\leq c}u_j
   \leq \sum_{1\leq j\leq c}(\hat{n}_j-\frac{1}{3}\hat{m}_j-1)
   \leq \sum_{1\leq j\leq c}(\hat{n}_j-\frac{1}{3}(\hat{n}_j-1)-1)
   \leq \frac{2}{3}n.
\end{displaymath}

Statement \ref{rel3}.  Let $n_i$ be the number of nodes of degree $i$
in $T$. Since $T$ is a tree of $n+1$ nodes, we have 
\begin{eqnarray*}
  2n &=&\sum_{i\geq 1}i\cdot n_i
        \geq n^*+n_2+\cdots+n_{k}+(k+1)\cdot N_k\label{eq:a};\\
  n+1&=& n^*+n_2+\cdots+n_{k}+N_k\label{eq:b}.
\end{eqnarray*}
$N_k\leq \frac{n}{k}$ follows immediately.
\end{proof}

\begin{lemma}\
\label{lemma:zzzz}

\begin{enumerate}
\item\label{item:first}
$\hat{G}_a$ has a weakly convenient encoding of
$2\hat{m}_a+2m_u+3n+o(\hat{m}_a+n)$ bits, from which the degree of a
node in $\hat{G}_a-T$ is obtainable in $O(1)$ time.

\item\label{item:second}
$\hat{G}_a$ has a convenient encoding of
$2\hat{m}_a+m_u+(4+\frac{1}{k})n+o(\hat{m}_a+n)$ bits, for any
positive constant $k$.
\end{enumerate}
\end{lemma}
\begin{proof}

Statement~\ref{item:first}.  Since each $T_j$ is a spanning tree of
$M_j$ that satisfies Lemma~\ref{lemma:canonical-tree}, $T$ is also a
spanning tree of $G_s$ that satisfies
Lemma~\ref{lemma:canonical-tree}.  Then, by
Lemma~\ref{lemma:weakly-convenient-simple-triconnected}, $G_a$ has a
weakly convenient encoding of $2m_a+2n+o(m_a+n)$ bits, from which the
degree of a node in $G_a-T$ is obtainable in $O(1)$ time. We next
extend this encoding to a desired weakly convenient encoding $X$ for
$\hat{G}_a$. Since $\hat{G}_a-T=G_a-T$, it suffices to add an $n$-bit
binary string $R$ such that $R[i]=1$ if and only if $v_i$ is
real. Since $m_a=\hat{m}_a+m_u$, the statement follows.

Statement~\ref{item:second}.  To augment the above encoding $X$ into a
convenient one for $\hat{G}_a$, it suffices to support in $O(1)$ time
a query on the number $r_i$ of real children of $v_i$ in $T$. Fix an
integer $k$.  Let $D$ be a binary string that contains $N_k$ copies of
$1$. If $v_i$ is the $h_i$-th node in $v_1,\ldots,v_n$ of degree more
than $k$ in $T$, we put $r_i$ copies of $0$ right after the $h_i$-th $1$ in
$D$. The length of $D$ is at most $N_k+n-m_u$.  Since $k=O(1)$, by the
definition of a weakly convenient encoding, it takes $O(1)$ time to
determine whether $d_i>k$ from $X$.  If $d_i\leq k$, $d_i$ and thus
the number of real neighbors of $v_i$ in $T$ can be computed in $O(1)$
time from $X$. If $d_i>k$, the number of real neighbors of $v_i$ in
$T$ is $\select{D,h_i+1,\mbox{1}}-\select{D,h_i,\mbox{1}}-1+R[i]$.  To
compute $h_i$ in $O(1)$ time, let $Y$ be an $n$-bit binary string such
that $Y[i]=1$ if and only if $d_i>k$.  Clearly if $d_i>k$, then
$h_i=\rank{Y,i,\mbox{1}}$, computable in $O(1)$ time from
$Y+\aux(Y)$. Since each $Y[i]$ can be determined in $O(1)$ time from
$X$, $Y$ need not be stored in our encoding. In summary, $X+D+\aux(D,Y)$
is a convenient encoding for $\hat{G}_a$, which can be coded in
$2\hat{m}_a+m_u+4n+N_k+o(\hat{m}_a+n)$ bits. The statement follows
immediately from Lemma~\ref{eq:mu-1}(\ref{rel3}).
\end{proof}

The next theorem summarizes the above discussion and extends
Lemma~\ref{lemma:zzzz} to accommodate multiple edges in $T$.
\begin{theorem}
\label{theorem:plane}
Let $\hat{G}$ be a plane graph of $n$ nodes and $\hat{m}$
edges. Assume that $\hat{G}_s$ is the simple version of $\hat{G}$.
\begin{enumerate}
\item\label{s1}
$\hat{G}$ $($respectively, $\hat{G}_s$$)$ has a weakly convenient
encoding of bit count $2\hat{m}+\frac{14}{3}n+o(\hat{m}+n)$ $($respectively,
$\frac{4}{3}\hat{m}_s+5n+o(n)$$)$.

\item\label{s2}
$\hat{G}$ $($respectively, $\hat{G}_s$$)$ has a convenient encoding of
$2\hat{m}+{\kfrac}n+o(\hat{m}+n)$ $($respectively,
$\frac{5}{3}\hat{m}_s+{\kfrac}n+o(n)$$)$ bits, for any positive
constant $k$.
\end{enumerate}
\end{theorem}
\begin{proof}
The statements for $\hat{G}_s$ follow immediately from Lemmas
\ref{eq:mu-1}(\ref{rel1}) and \ref{lemma:zzzz} with
$\hat{G}_a=\hat{G}_s$.  To prove the statements for $\hat{G}$, we
first choose $E_a$ to be the set of multiple edges such that $(G_s-T)
\cup E_a$ is composed of the multiple edges of $\hat{G}$ between
adjacent nodes in $G_s-T$.  Also, let $E_b=\hat{G}-\hat{G}_a$; let
$\ell_b$ be the number of edges in $E_b$.

Statement~\ref{s1}.  Continuing the proof of
Lemma~\ref{lemma:zzzz}(\ref{item:first}), we augment the weakly
convenient encoding $X$ for $\hat{G}_a$ into one for $\hat{G}$. We
support in $O(1)$ time a query for the number $a_i$ of multiple edges
of $\hat{G}$ between $v_i$ and its parent in $T$ as follows.

Initially, $L_0$ is $n-c$ copies of $1$, one for each node that is not
in the first two levels of $T$; recall that all nodes in the first two
levels of $T$ are unreal. For $1\leq i\leq n$, suppose that $v_i$ is
the $g_i$-th node in $v_1,\ldots,v_n$ that is not in the first two
levels of $T$. We put $a_i$ copies of $0$ right after the $g_i$-th $1$
in $L_0$. Since $\hat{G}$ has $n+\ell_b-m_u$ edges between adjacent
nodes in $T$, $L_0$ has $2n-c+\ell_b-m_u$ bits.

Let $L$ be an $n$-bit binary string such that for $1\leq i\leq n$,
$L[i]=1$ if and only if $v_i$ is not in the first two levels of
$T$. Clearly if $L[i]=1$, then $g_i=\rank{L,i,1}$. Since $L[i]$ is
obtainable from $X$ in $O(1)$ time, $a_i$ is obtainable in $O(1)$ time
from $X+L_0+\aux(L_0,L)$. Moreover, since $R[i]=1$ if and only
$a_i\geq 1$, $R$ can be removed from $X$. Thus, $\hat{G}$ has a weakly
convenient encoding $\hat{X}$ of
$2\hat{m}_a+m_u+4n-c+\ell_b+o(\hat{m}_a+n)$ bits. The statement
follows from Lemma~\ref{eq:mu-1}(\ref{rel2}) and the fact that
$\hat{m}=\hat{m}_a+\ell_b$.

Statement~\ref{s2}.  We now augment the above encoding $\hat{X}$ into a
convenient one for $\hat{G}$.  It suffices to support in $O(1)$ time a
query on the number $r_i$ of the real multiple edges $\hat{G}$ between
$v_i$ and its children in $T$. Initially, $D$ is $N_k$ copies of
$1$. Suppose that $v_i$ is the $h_i$-th node in $v_1,\ldots,v_n$ of
degree more than $k$ in $T$. We put $r_i$ copies of $0$ right after the
$h_i$-th $1$ in $D$. As in the proof of
Lemma~\ref{lemma:zzzz}(\ref{item:second}), $h_i$ is obtainable from
$Y+\aux(Y)$ in $O(1)$ time, where $Y$ is not stored in the encoding. If
$d_i>k$, $r_i$ is computable as
$\select{D,h_i+1,\mbox{1}}-\select{D,h_i,\mbox{1}}-1$ in $O(1)$ time. If
$d_i\leq k$, $r_i$ is computable in $O(k)$ time from $\hat{X}$. $D$ has
at most $N_k+n+\ell_b-m_u$ bits. Hence $G$ has a convenient encoding
$\hat{X}+D+\aux(D,Y)$ of
$2\hat{m}_a+5n+2\ell_b+N_k-c+o(\hat{m}_a+\ell_b+n)$ bits. Then, this
statement follows from Lemma~\ref{eq:mu-1}(\ref{rel3}) and the fact
$\hat{m}=\hat{m}_a+\ell_b$.
\end{proof}

\subsection{Graphs with Self-loops}\
\label{sec:self-loops} 

{\it Remark.}  The encodings of Theorems
\ref{theorem:convenient-triconnected},
\ref{theorem:triangulated}, and \ref{theorem:plane}
assume that $G$ has no self-loops.  To facilitate the coding of
self-loops, we assume that the self-loops incident to a node in a
plane graph are recorded at that node by their number.  Then, to
augment each cited encoding to accommodate self-loops, we only need
to add $1$ to the coefficient of the term $n$ in the bit count as
follows.  Initially, $Z$ is $n$ copies of $1$. Then, for $1 \leq i
\leq n$, we put $z_i$ copies of $0$ right after the $i$-th $1$ in $Z$,
where $z_i$ is the number of self-loops incident to $v_i$. We augment
the encoding in question with $Z$ by means of Fact~\ref{fact_aux}.
Since the bit count of $Z$ is $n$ plus the number of self-loops, our
claims follows from the fact that the coefficient of the term $m$ in
the bit count in question is at least one.

\section{More Compact Schemes}\label{compact_coding} 
For applications that require no query support, we obtain more compact
encodings for triconnected plane graphs in this section.
All graphs in this section are simple.

% Comment out the old Subsection 4.1

% The following heading for the old Subsection 4.2 is commented out
% \subsection{Triconnected Plane Graphs}\label{sec_triconnect} 
Let $G$
be a triconnected plane graph with $n > 3$ nodes. Let $T$ be a
canonical spanning tree of $G$. Let $v_1,\ldots,v_n$ be the
counterclockwise preordering of $T$, which by
Fact~\ref{fact:preordering} is also a canonical ordering of $G$.

Let $I_1,\ldots,I_K$ be the interval partition for the ordering
$v_1,\ldots,v_n$.  Recall that the construction algorithm of
\S\ref{section:canonical} builds $G$ from a single edge $(v_1,v_2)$
through a sequence of $K$ steps.  The $j$-th step corresponds to the
interval $I_j=[k,k+q]$. There are two cases, which are used throughout
this section.

{\em Case} 1: $q=0$, and a single node $v_k$ is added.

{\em Case} 2: $q > 0$, and a chain of $q+1$ nodes $v_k,\ldots,v_{k+q}$
is added.

The last node added during a step is called {\em type a}; the other
nodes are {\em type b}. Thus for a Case 1 step, $v_k$ is type a. For a
Case 2 step, $v_k,v_{k+1},\ldots,v_{k+q-1}$ are type b, and the node
$v_{k+q}$ is type a.  To define further terms, let $c_1
=v_1,c_2,\ldots,c_t =v_2$ be the nodes of $H_{k-1}$ ordered
consecutively along $H_{k-1}$ from left to right above the edge
$(v_1,v_2)$.

Case 1. Let $c_\ell$ and $c_r$, where $1\leq \ell < r \leq t$, be the
leftmost and rightmost neighbors of $v_k$ in $H_{k-1}$,
respectively. The edge $(c_r, v_k)$ is called {\em external}. The
edges $(c_i,v_k)$ where $\ell < i < r$, if present, are {\em
internal}. Note that $(c_\ell,v_k)$ is in $T$.

Case 2. Let $c_\ell$ and $c_r$, where $1\leq \ell < r \leq t$, be the
neighbors of $v_k$ and $v_{k+q}$ in $H_{k-1}$, respectively.  The edge
$(c_r,v_k)$ is called {\it external}.  Observe that the edges
$(c_\ell,v_k),(v_k,v_{k+1}),\ldots$, $(v_{k+q-1},v_{k+q})$ are in $T$.

For each $v_h$, where $1\leq h \leq n-1$, let $B(v_h)$ denote the edge
set $\{ (v_h,v_j)~|~h < j\}$. By the definition of a canonical
ordering and Lemma~\ref{lemma:canonical-tree}, the edges in $B(v_h)$
form the following pattern around $v_h$ in the counterclockwise order:
a block (maybe empty) of tree edges; followed by at most one internal
edge; followed by a block (maybe empty) of external edges. Note that
$B(v_1),B(v_2),\ldots,B(v_{n-1})$ form a partition of the edges of
$G$. Also, $B(v_h)$ is not empty since by the definition of a
canonical ordering, every $v_h$ is adjacent to some $v_j$ with $h <
j$.

\begin{lemma}\label{lemma:reconstruct2}
Given $B(v_h)$ for $1\leq h \leq n-1$ and the type of $v_h$ for $3
\leq h \leq n$, we can uniquely reconstruct $G$.
\end{lemma}
\begin{proof}
We first draw $(v_1,v_2)$ and then perform the following $K$
steps. Step $j$ processes $I_j=[k,k+q]$. Before this step, $G_{k-1}$
and $H_{k-1}$ have been built. Let $c_1 =v_1,c_2,\ldots, c_t =v_2$ be
the nodes on $H_{k-1}$ from left to right. We know the numbers of {\it
remaining} tree and external edges at each $c_i$, i.e., those in
$B(c_i)$ not yet added to $G$.  We next find the leftmost neighbor
$c_{\ell}$ and the rightmost neighbor $c_r$ of the nodes added during
this step. Note that $(c_{\ell},v_k)$ is in $T$.  Since
$v_1,\ldots,v_n$ is the counterclockwise preordering of $T$, $c_\ell$
is the rightmost node with a remaining tree edge; $c_r$ is the
leftmost node to the right of $c_\ell$ with a remaining external
edge. There are two cases:

If $v_k$ is type a, then this is a Case 1 step and $v_k$ is the single
node added during this step. We add $(c_\ell,v_k)$ and
$(c_r,v_k)$. For each $c_i$ with $\ell < i < r$, if $B(c_i)$ contains
an internal edge, we also add $(c_i,v_k)$.

If $v_k$ is type b, then this is a Case 2 step.
Let $q$ be the integer such that $v_k,v_{k+1},\ldots,$
$v_{k+q-1}$ are type b and $v_{k+q}$ is type a. The chain
$v_k,\ldots,v_{k+q}$ is added between $c_\ell$ and $c_r$.

Finally, the number of remaining tree (respectively, external) edges
at $c_{\ell}$ (respectively, $c_r$) decreases by 1. The numbers of
tree, internal, and external edges remaining at each $v_i$ for $k \leq
i \leq k+q$ are set to those of all tree, internal, and external edges
in $B(v_i)$. This finishes the $j$-th step. When the $K$-th step ends,
we have $G$.
\end{proof}

By Lemma~\ref{lemma:reconstruct2}, we can encode $G$ by encoding the
types of all $v_h$ and $B(v_h)$ for $1\leq h\leq n-1$ using two
strings $S_1$ and $S_2$. $S_1$ is a binary string containing one bit
for each $v_h$, indicating the type of $v_h$. $S_2$ encodes the sets
$B(v_h)$ using three symbols $0,1,*$. The code for $B(v_h)$ is a block
of 0's, followed by a block of 1's, followed by a block of $*$'s.  The
number of 0's (respectively, 1's and $*$'s) in the first
(respectively, second and third) block is that of the tree
(respectively, external and internal) edges in $B(v_h)$. However,
since these three numbers can be zero, we need a fourth symbol to
separate the codes for $B(v_h)$. Now if we use two bits to encode each
of the 4 symbols used in $S_2$, then $S_2$ has a longer binary
encoding than desired.  We next present a shorter encoding by
eliminating the symbol used to separate the codes for $B(v_h)$.

The {\em type} of $B(v_h)$ is defined to be a combination of symbols
$T,X$ and $I$, which denote the existences of tree, external or
internal edges in $B(v_h)$, respectively. For example, if $B(v_h)$ is
type $TI$, then it has at least one tree edge, exactly one internal
edge, and no external edge; recall that each $B(v_h)$ has at most one
internal edge.  Moreover, for all $v_h$ of type a, if $B(v_h)$ has no
tree edge, then we call $v_h$ {\em type a1}; otherwise, $v_h$ is {\em
type a2}.  For $v_h$ of type b, since $v_h$ is added in a Case 2 step
and is not the last node added, $B(v_h)$ has at least one tree edge
and thus no similar typing is needed.

Our encoding of $G$ uses two strings $S_1$ and $S_2$. $S_1$ has length
$n$. For $1\leq h \leq n$, $S_1[h]$ indicates whether $v_h$ is type
a1, a2, or b, which is recorded by symbols $0,1$, or $*$,
respectively.  For convenience, let $v_1$ be type a2 and $v_2$ be type
a1.  $S_2$ uses the same three symbols to encode $B(v_h)$ for $1\leq h
\leq n-1$.  $B(v_h)$ is specified by a codeword $\code{v_h}$ defined
in Figure \ref{table}.  $S_2$ is the concatenation of the codewords
$\code{v_h}$.

\begin{figure}
\begin{center}
\begin{tabular}{||l|l|l||} \hline
type of $v_h$ & type of $B(v_h)$ & $\code{v_h}$     \\ \hline
a1      & $XI$     & $\underbrace{1^{\beta}}_X\underbrace{0}_I$\\ \cline{2-3}
        & $I$      & $\underbrace{0}_I$                        \\ \cline{2-3}
        & $X$      & $\underbrace{1^{\beta-1}*}_X$             \\ \hline
a2 or b & $T$      & $\underbrace{0^{\alpha -1}*}_T$  \\ \cline{2-3}
        & $TXI$    & $\underbrace{1^{\alpha}}_T\underbrace{0^{\beta}}_X
\underbrace{*}_I$  \\ \cline{2-3}
        & $TX$     & $\underbrace{1^{\alpha-1}0}_T
\underbrace{0^{\beta -1}1}_X$ \\ \cline{2-3}
        & $TI$     & $\underbrace{1^{\alpha}}_T\underbrace{*}_I$ \\ \hline
\end{tabular}
\caption{This 
code book gives $\code{v_h}$.  The length of $\code{v_h}$ is the number
of edges in $B(v_h)$.  The numbers of the tree and external edges in
$B(v_h)$ are denoted by $\alpha$ and $\beta$, respectively.  Recall
that $B(v_h)$ contain either 0 or 1 internal edge. The notation $z^t$
denotes a string of $t$ copies of symbol $z$. A symbol $T$, $X$, or
$I$ under $\code{v_h}$ denotes the portion in $\code{v_h}$
corresponding to the tree, external, or internal edges, respectively.}
\label{table}
\end{center}
\end{figure}

\begin{lemma}\label{lemma:reconstructS} For $1\leq h\leq n-1$, the
sets $B(v_h)$ and the types of all $v_h$ can be uniquely determined
from $S_1$ and $S_2$.
\end{lemma}
\begin{proof}
We can look up the type of $v_h$ in $S_1$. To recover $B(v_h)$, we
perform the following $n-1$ steps. Before step $h$, we know the start
index of $\code{v_h}$ in $S_2$.  With the cases below, step $h$ finds
the numbers of tree, external, and internal edges in $B(v_h)$ as well
as the length of $\code{v_h}$, which tells us the start index of
$\code{v_{h+1}}$ in $S_2$.

{\it Case A}: $v_h$ is type a1. There are three subcases.

{\it Case A1}: $\code{v_h}$ starts with 0. Then $B(v_h)$ is type $I$
and contains only one internal edge. Also, $\code{v_h}$ has length 1.

{\it Case A2}: $\code{v_h}$ starts with $*$. Then $B(v_h)$ is type $X$
with $\beta =1$ external edge. Also, $\code{v_h}$ has length 1.

{\it Case A3}: $\code{v_h}$ starts with 1. Let $\Theta=1^\gamma$ be
the maximal block of 1's in $S_2$ at the start of $\code{v_h}$.  Then,
$\code{v_h}$ has length $\gamma+1$.  Let $x$ be the symbol after
$\Theta$ in $S_2$. There are two further subcases.

If $x = *$, $B(v_h)$ is type $X$ and has $\beta=\gamma+1$ external
edges.

If $x = 0$, $B(v_h)$ is type $XI$ and has $\beta=\gamma$ external
edges and one internal edge.

{\it Case B}: $v_h$ is type a2 or b. Then $B(v_h)$ contains at least
one tree edge. There are three subcases.

{\it Case B1}: $\code{v_h}$ starts with $*$. Then $B(v_h)$ is type $T$
and contains $\alpha =1$ tree edge. Also, $\code{v_h}$ has length 1.

{\it Case B2}: $\code{v_h}$ starts with 0. Let $\Theta=0^\gamma$ be
the maximal block of 0's in $S_2$ at the start of $\code{v_h}$.  Then
$\code{v_h}$ has length $\gamma+1$.  Let $x$ be the symbol after
$\Theta$ in $S_2$. There are two further subcases:

If $x=*$, then $B(v_h)$ is type $T$ and has $\alpha=\gamma+1$
tree edges.

If $x= 1$, then $B(v_h)$ is type $TX$ and has 1 tree edge and
$\beta=\gamma$ external edges.

{\it Case B3}: $\code{v_h}$ starts with 1. Let $\Theta = 1^\gamma$ be
the maximal block of 1's in $S_2$ at the start of $\code{v_h}$.  There
are three further subcases:

If $*$ follows $\Theta$ in $S_2$, then $B(v_h)$ is type $TI$ and has
$\alpha = \gamma$ tree edges and one internal edge. Also, $\code{v_h}$
has length $\gamma+1$.

If $0^\delta*$ follows $\Theta$ in $S_2$, then $B(v_h)$ is type $TXI$
and has $\alpha=\gamma$ tree edges, $\beta=\delta$ external edges, and
one internal edge.  Also, $\code{v_h}$ has length $\gamma+\delta+1$.

If $0^\delta 1$ follows $\Theta$ in $S_2$, then $B(v_h)$ is type $TX$
and has $\alpha=\gamma+1$ tree edges and $\beta=\delta$ external
edges.  Also, $\code{v_h}$ has length $\gamma+\delta+1$.

This completes the description of the $h$-th step.  In any case above,
we can determine the length of $\code{v_h}$ and recover $B(v_h)$.
\end{proof}

The next theorem summarizes the above discussion.
\begin{theorem}
Let $G$ be a simple triconnected plane graph with $n > 3$ nodes, $m$
edges, and $f$ faces.
\begin{enumerate}
\item 
$G$ can be encoded using at most $\log 3{\cdot}(n+m) +1$ bits.
\item 
$G$ can be encoded using at most $\log 3{\cdot}(\min\{n,f\}+m)+2 \leq
\frac{3\log 3}{2}m+4$ bits.
\end{enumerate}
\end{theorem}

{\it Remark.}  The decoding procedure assumes that the encoding of $G$
is given together with $n$ or $f$ as appropriate, which can be
appended to $S$ by means of Fact~\ref{fact_aux}.

\begin{proof}

Statement 1.  In the above discussion, $S_1$ has length $n$, and $S_2$
has length $m$. The encoding $S$ of $G$ is the concatenation of $S_1$
and $S_2$. Treated as an integer of base 3, $S$ uses at most $\log
3{\cdot}(n+m) + 1$ bits.

Statement 2. Let $G^*$ be the dual of $G$. $G^*$ has $f$ nodes, $m$
edges and $n$ faces. Since $G$ is triconnected, $G^*$ is also
triconnected. Furthermore, since $n > 3$, $f > 3$ and $G^*$ has no
self-loop or multiple edge. Thus, we can use Statement 1 to encode
$G^*$ with at most $\log 3{\cdot}(f+m) +1$ bits. Since $G$ can be uniquely
determined from $G^*$, to encode $G$, it suffices to encode $G^*$. To
shorten $S$, if $n \leq f$, we encode $G$ using at most $\log 3{\cdot}(n+m)
+1$ bits; otherwise, we encode $G^*$ using at most $\log
3{\cdot}(f+m)+1$ bits. This new encoding uses at most $\log
3{\cdot}(\min\{n,f\}+m)+1$ bits. Since $\min\{n,f\} \leq
\frac{n+f}{2}=0.5m+1$, the bit count is at most $\log
3{\cdot}(1.5m)+3$. For the sake of decoding, we use one extra bit to
denote whether we encode $G$ or $G^*$.
\end{proof}


\begin{thebibliography}{10}

\bibitem{ahu74}
{\sc A.~V. Aho, J.~E. Hopcroft, and J.~D. Ullman}, {\em The Design and Analysis
  of Computer Algorithms}, Addison-Wesley, Reading, MA, 1974.

\bibitem{BCW90}
{\sc T.~C. Bell, J.~G. Cleary, and I.~H. Witten}, {\em Text Compression},
  Prentice-Hall, Englewood Cliffs, NJ, 1990.

\bibitem{Ber85}
{\sc C.~Berge}, {\em Graphs}, North-Holland, New York, NY, second revised~ed.,
  1985.

\bibitem{Clark96}
{\sc D.~R. Clark}, {\em Compact Pat Trees}, PhD thesis, University of Waterloo,
  1996.

\bibitem{DeFPP90}
{\sc H.~de~Fraysseix, J.~Pach, and R.~Pollack}, {\em How to draw a planar graph
  on a grid}, Combinatorica, 10 (1990), pp.~41--51.

\bibitem{Elias75}
{\sc P.~Elias}, {\em Universal codeword sets and representations of the
  integers}, IEEE Transactions on Information Theory, IT-21 (1975),
  pp.~194--203.

\bibitem{GW83}
{\sc H.~Galperin and A.~Wigderson}, {\em Succinct representations of graphs},
  Information and Control, 56 (1983), pp.~183--198.

\bibitem{GrotschelLS88}
{\sc M.~Gr{\"{o}}tschel, L.~Lov\'{a}sz, and A.~Schrijver}, {\em Geometric
  Algorithms and Combinatorial Optimization}, Springer-Verlag, New York, NY,
  1988.

\bibitem{hkaol.coding_one}
{\sc X.~He, M.~Y. Kao, and H.~I. Lu}, {\em Linear-time succinct encodings of
  planar graphs via canonical orderings}, {SIAM} Journal on Discrete
  Mathematics,  (1999).
\newblock To appear.

\bibitem{IR82}
{\sc A.~Itai and M.~Rodeh}, {\em Representation of graphs}, Acta Informatica,
  17 (1982), pp.~215--219.

\bibitem{Jacobson89}
{\sc G.~Jacobson}, {\em Space-efficient static trees and graphs}, in
  Proceedings of the 30th Annual IEEE Symposium on Foundations of Computer
  Science, 1989, pp.~549--554.

\bibitem{KNR92}
{\sc S.~Kannan, M.~Naor, and S.~Rudich}, {\em Implicit representation of
  graphs}, {SIAM} Journal on Discrete Mathematics, 5 (1992), pp.~596--603.

\bibitem{Kant92}
{\sc G.~Kant}, {\em Drawing planar graphs using the $lmc$-ordering}, in
  Proceedings of the 33rd Annual IEEE Symposium on Foundations of Computer
  Science, 1992, pp.~101--110.

\bibitem{Kant93}
\leavevmode\vrule height 2pt depth -1.6pt width 23pt, {\em Algorithms for
  Drawing Planar Graphs}, PhD thesis, University of Utrecht, 1993.

\bibitem{KH97}
{\sc G.~Kant and X.~He}, {\em Regular edge labeling of 4-connected plane graphs
  and its applications in graph drawing problems}, Theoretical Computer
  Science, 172 (1997), pp.~175--193.

\bibitem{kaofhr94}
{\sc M.~Y. Kao, M.~F\"urer, X.~He, and B.~Raghavachari}, {\em Optimal parallel
  algorithms for straight-line grid embeddings of planar graphs}, {SIAM}
  Journal on Discrete Mathematics, 7 (1994), pp.~632--646.

\bibitem{kaot93.joa}
{\sc M.~Y. Kao, N.~Occhiogrosso, and S.~H. Teng}, {\em Simple and efficient
  compression schemes for dense and complement graphs}, Journal of
  Combinatorial Optimization, 2 (1999), pp.~351--359.

\bibitem{KW:encodings}
{\sc K.~Keeler and J.~Westbrook}, {\em Short encodings of planar graphs and
  maps}, Discrete Applied Mathematics, 58 (1995), pp.~239--252.

\bibitem{Munro96}
{\sc J.~I. Munro}, {\em Tables}, in Lecture Notes in Computer Science 1180:
  Proceedings of the 16th Conference on Foundations of Software Technology and
  Theoretical Computer Science, Springer-Verlag, New York, NY, 1996,
  pp.~37--42.

\bibitem{MR97}
{\sc J.~I. Munro and V.~Raman}, {\em Succinct representation of balanced
  parentheses, static trees and planar graphs}, in Proceedings of the 38th
  Annual IEEE Symposium on Foundations of Computer Science, 1997, pp.~118--126.

\bibitem{naor90}
{\sc M.~Naor}, {\em Succinct representations of general unlabeled graphs},
  Discrete Applied Mathematics, 28 (1990), pp.~303--307.

\bibitem{PH86.encode}
{\sc C.~H. Papadimitriou and M.~Yannakakis}, {\em A note on succinct
  representations of graphs}, Information and Control, 71 (1986), pp.~181--185.

\bibitem{Read87}
{\sc R.~C. Read}, {\em A new method for drawing a planar graph given the cyclic
  order of the edges at each vertex}, Congressus Numerantium, 56 (1987),
  pp.~31--44.

\bibitem{Schnyder90}
{\sc W.~Schnyder}, {\em Embedding planar graphs on the grid}, in Proceedings of
  the 1st Annual ACM-SIAM Symposium on Discrete Algorithms, 1990, pp.~138--148.

\bibitem{turan84}
{\sc G.~Tur\'{a}n}, {\em On the succinct representation of graphs}, Discrete
  Applied Mathematics, 8 (1984), pp.~289--294.

\bibitem{Tutte62}
{\sc W.~T. Tutte}, {\em A census of planar triangulations}, Canadian Journal of
  Mathematics, 14 (1962), pp.~21--38.

\end{thebibliography}
\end{document}